\documentclass[twocolumn,twocolappendix]{aastex701}

\usepackage{graphicx}
\usepackage{array}
\usepackage{multirow}
\usepackage[sort&compress]{natbib}
\usepackage{hyperref}
\usepackage{rotating}
\usepackage{graphicx}
\usepackage{longtable}
\usepackage{CJK}
\usepackage[T1]{fontenc}
\usepackage[utf8]{inputenc}
\usepackage{tikz}
\usepackage{threeparttable}
\usepackage{gensymb}
\usepackage{soul}
\usepackage{enumitem}

\usepackage{amsmath}
\usepackage{booktabs}
\usepackage{array}

\begin{document}
\begin{CJK*}{UTF8}{gbsn}
\title{Stellar Loci XI: Identifications and Parameter Determinations of Binary Stars by multi-band photometry}
\correspondingauthor{Haibo Yuan}
\email{yuanhb@bnu.edu.cn}

\author[0009-0000-7044-4393]{Yuanchang Wang (王源畅)}
\affiliation{Institute for Frontiers in Astronomy and Astrophysics,
Beijing Normal University, Beijing, 102206, China}
\affiliation{School of Physics and Astronomy, Beijing Normal University No.19, Xinjiekouwai St, Haidian District, Beijing, 100875, China}
\email{wangyc@mail.bnu.edu.cn}

\author[0000-0003-2471-2363]{Haibo Yuan (苑海波)}
\affiliation{Institute for Frontiers in Astronomy and Astrophysics,
Beijing Normal University, Beijing, 102206, China}
\affiliation{School of Physics and Astronomy, Beijing Normal University No.19, Xinjiekouwai St, Haidian District, Beijing, 100875, China}
\email{yuanhb@bnu.edu.cn}

\author[0000-0002-3651-0681]{Zexi Niu (牛泽茜)}
\affiliation{School of Astronomy and Space Science, University of Chinese Academy of Sciences, Beijing 100049, China}
\affiliation{National Astronomical Observatories, Chinese Academy of Sciences 20A Datun Road, Chaoyang District Beijing 100101, China}
\email{nzx@nao.cas.cn}

\author[0000-0001-8424-1079]{Kai Xiao (肖凯)}
\affiliation{National Astronomical Observatories, Chinese Academy of Sciences 20A Datun Road, Chaoyang District Beijing 100101, China}
\email{xiaokai@bao.ac.cn}

\author[0000-0002-1259-0517]{Bowen Huang ({黄博闻})}
\affiliation{National Astronomical Observatories, Chinese Academy of Sciences 20A Datun Road, Chaoyang District Beijing 100101, China}
\affiliation{School of Physics and Astronomy, Beijing Normal University No.19, Xinjiekouwai St, Haidian District, Beijing, 100875, China}
\affiliation{Institute for Frontiers in Astronomy and Astrophysics, Beijing Normal University, Beijing, 102206, China}
\email[]{huangbw@bao.ac.cn}

\author[0009-0000-0602-0801]{Ziqi Gao (高子琪)}
\affiliation{Institute for Frontiers in Astronomy and Astrophysics,
Beijing Normal University, Beijing, 102206, China}
\affiliation{School of Physics and Astronomy, Beijing Normal University No.19, Xinjiekouwai St, Haidian District, Beijing, 100875, China}
\email{gziqi@mail.bnu.edu.cn}

\author[0000-0002-5818-8769]{Maosheng Xiang (向茂盛)}
\affiliation{National Astronomical Observatories, Chinese Academy of Sciences 20A Datun Road, Chaoyang District Beijing 100101, China}
\affiliation{Institute for Frontiers in Astronomy and Astrophysics,
Beijing Normal University, Beijing, 102206, China}
\email{msxiang@nao.cas.cn}

\begin{abstract}

Large photometric surveys contain numerous unresolved binaries, but identifying them and measuring component properties remains challenging. We present a 65-band UV-to-IR SED-fitting framework to identify unresolved main-sequence binaries and infer their parameters. Photometry combines synthetic miniJPAS magnitudes from corrected Gaia DR3 XP spectra with 2MASS and WISE measurements. Using LAMOST-calibrated, metallicity-dependent stellar loci and a dwarf-star color--absolute-magnitude relation, we construct forward models for single stars and unresolved binaries. We fit both model families to each SED with Markov chain Monte Carlo (MCMC), using Gaia distances as Hertzsprung--Russell-diagram constraints to infer metallicity, reddening, and component photometric parameters. Component masses and mass ratios are derived from PARSEC models.
Simulations show that the method robustly recovers binary parameters. Tests on a LAMOST comparison sample indicate sensitivity to binaries with $q\gtrsim0.4$; the preferred model reproduces spectroscopic metallicities with 0.11 dex scatter, while inferred reddenings agree with external extinction maps with $\sim0.02$ mag scatter.
Comparison with an external Gaia XP--based binary catalog shows broadly consistent binary classifications for common sources. Our metallicities more closely follow the LAMOST spectroscopic scale, and mass-ratio estimates in the intermediate-$q$ regime better trace expected $\Delta$-HRD offsets. Tests using M67 members support internal consistency of inferred parameters. For quality-selected Gaia non-single-star binaries, primary-mass estimates agree with Gaia NSS values with 0.071 \(M_\odot\) residual scatter; in cases with larger discrepancies, further tests indicate that secondary-mass and mass-ratio estimates are more reliable than the Gaia NSS values. The method scales to large photometric samples, enabling population studies and analyses of individual systems.

\end{abstract}


\section{Introduction} \label{sec:intro}

Binary stars are common throughout the Universe; in the Milky Way, nearly half of all stars are thought to reside in binary systems \citep{2013ARA&A..51..269D, 2017ApJS..230...15M}. Interactions in such systems shape the formation and evolution of a wide range of astrophysical phenomena, including Type Ia supernovae (e.g., \citealt{2000ApJS..128..615M}), short-duration gamma-ray bursts (e.g., \citealt{1992ApJ...395L..83N}), kilonovae (e.g., \citealt{2017LRR....20....3M}), cataclysmic variables (e.g., \citealt{1995cvs..book.....W}), X-ray binaries (e.g., \citealt{2024ApJ...971..133R}), stellar population synthesis (e.g., \citealt{2002MNRAS.329..897H}), and gravitational-wave sources (e.g., \citealt{2017PhRvL.119p1101A}).
Key orbital parameters—mass ratio, period, eccentricity, and component masses—encode the formation and evolutionary history of binary systems (\citealt{2020RAA....20..161H}). Accordingly, a variety of techniques have been developed to infer binary orbits and physical properties, including visual-binary measurements (e.g., \citealt{1996AJ....111..370H}), radial-velocity (RV) monitoring (e.g., \citealt{2025ApJS..278...46G}), eclipsing-binary analyses (e.g., \citealt{2019MNRAS.485..961J}), and the identification of co-moving pairs using astrometric data (e.g., \citealt{2019MNRAS.482L.139E}).

Although these direct techniques can be highly accurate, they typically require multi-epoch observations and are subject to strong selection effects tied to orbital inclination, period, and mass ratio (\citealt{2016BaltA..25..384M}; \citealt{2018MNRAS.476..528E}). As a result, available samples remain relatively small, limiting constraints on the global properties of binary-star populations. Motivated by the advent of large-scale surveys, a number of indirect methods have therefore been developed to study binary systems.

To mitigate these biases and limitations, \cite{2015ApJ...799..135Y} introduced the stellar locus outlier (SLOT) method, which enables model-independent estimates of the binary fraction for large stellar samples spanning diverse populations and wide survey volumes. Applying SLOT to SDSS Stripe 82 and LAMOST samples, \cite{2015ApJ...799..135Y} measured binary fractions across stellar populations and investigated trends with spectral type and metallicity ([Fe/H]), finding a mild but statistically significant decrease in binary fraction with increasing [Fe/H]. Using Gaia Early Data Release 3 together with LAMOST Data Release 5, \cite{2021ApJ...922..211N} obtained bias-corrected binary fractions for field late-G and early-K dwarfs and reported a positive correlation between [$\alpha$/Fe] and binary fraction for thin-disk stars at fixed metallicity; this trend was not observed for thick-disk stars.

Main-sequence binaries also form a sequence brighter than single stars in the Hertzsprung--Russell diagram (HRD), motivating a range of HRD-based statistical approaches to binary populations (\citealt{2019MNRAS.490..550L}; \citealt{2020ApJ...901...49L}; \citealt{2025ApJS..277...15J}).
Nevertheless, most such indirect methods have remained inherently statistical because earlier photometric datasets---limited by filter bandwidth, spectral coverage, and measurement uncertainties---could not deliver precise constraints for individual sources.

The limited availability of binaries with well-measured parameters---a legacy of historical data quality and detection constraints---has left key aspects of binary populations uncertain, as reflected in ongoing debates over the binary fraction and the form of the mass-ratio distribution (\citealt{2003ApJ...591..397G}; \citealt{2013ARA&A..51..269D}). Accordingly, developing methods that can deliver precise parameters for large binary samples remains a central goal of modern astrophysical research.

The availability of Gaia low-resolution $BP/RP$ (hereafter XP) spectrophotometry, together with miniJPAS's 60--filter photometry, opens a new avenue for identifying and characterizing unresolved main-sequence binaries. \citet{2025A&A...704A.126L} showed that Gaia XP spectra combined with Gaia parallaxes can be used to detect such systems by fitting composite binary models---constructed by superposing the absolute fluxes of the primary and secondary components---and comparing the goodness of fit of single-star and binary-star models. Using this approach, they compiled a largely orbit-independent, model-based census of $\sim$14 million main-sequence binary candidates within 1 kpc and inferred key parameters including primary mass, photometric metallicity, and mass ratio. Their method, however, is less sensitive to low mass-ratio binaries ($q \lesssim 0.5$) and struggles to distinguish systems with very faint companions.

As shown by \citet{2015ApJ...799..135Y}, unresolved binaries produce wavelength-dependent color offsets relative to single-star loci, and the diagnostic power of these offsets varies across photometric bands. With its contiguous narrow-band coverage, miniJPAS can resolve these spectral-shape signatures in detail. When coupled with the luminosity offset of binaries in the HRD, this multi-band information offers a path to improving upon earlier approaches and enabling precise parameter inference for both large samples and lower mass-ratio systems.

In this work, we convolve Gaia XP spectra with the miniJPAS filter response curves to derive magnitudes in 60 bands. We further incorporate infrared photometry from 2MASS (\(J\), \(H\), \(Ks\)) and WISE (\(W1\), \(W2\)), and construct 64 metallicity-dependent stellar loci together with a metallicity-dependent color--absolute-magnitude relation. These empirical relations enable tight constraints on individual binaries, including metallicity and the colors and magnitudes of both components.

This paper is organized as follows. Section~\ref{sec:DATA} describes the observational datasets and selection criteria, with an emphasis on AFGKM dwarfs. Section~\ref{sec:METHODS} outlines the single- and binary-star models and the empirical relations used in our analysis. Section~\ref{sec:RESULTS AND DISCUSSION} presents the results and compares them with previous work. Section~\ref{sec:SUMMARY} summarizes the main findings and discusses directions for future research.

\section{DATA} \label{sec:DATA}

\subsection{Gaia DR3 Data}\label{subsec:Gaia}

The ESA \,\textit{Gaia} mission provides astrometry, photometry, and spectroscopy for sources across the Milky Way and Local Group (\citealt{2016A&A...595A...1G}). Gaia DR3 contains photometry for over 1.8 billion sources and five-parameter astrometry (sky position, parallax, and proper motions) for 1.47 billion sources (\citealt{2023A&A...674A...1G}), as well as $\sim$220 million XP spectra spanning 336--1020\,nm (\citealt{2023A&A...674A...1G}; \citealt{2023A&A...674A...2D}). We apply the \cite{2024ApJS..271...13H} corrections to the XP spectra to mitigate magnitude-, color-, and extinction-dependent systematics, achieving an internal precision of 1--2\%.

\subsection{LAMOST} \label{subsec:LAMOST}

LAMOST, operated by the National Astronomical Observatory of China, is a 4\,m quasi-meridian reflective Schmidt telescope equipped with 4000 fibers. Data Release 10 (DR10) provides 11,817,430 low-resolution spectra spanning 369--910\,nm at a resolving power of $R\sim1800$. The LAMOST Stellar Parameter Pipeline (\citealt{2011RAA....11..924W}; \citealt{2015RAA....15.1095L}) derives $T_{\mathrm{eff}}$, $\log g$, and $\mathrm{[Fe/H]}$ for late-A and FGK stars. For spectra with signal-to-noise ratios $>20$, typical internal uncertainties are $\sim$50--100\,K in $T_{\mathrm{eff}}$, $\sim$0.05--0.1\,dex in $\log g$, and $\sim$0.05--0.1\,dex in $\mathrm{[Fe/H]}$. Using wide binaries, \cite{2023ApJ...950..104N} quantified systematic abundance offsets for FGK dwarfs (their Figure~4), finding that LAMOST $\mathrm{[Fe/H]}$ can be underestimated by up to 0.4\,dex at $T_{\mathrm{eff}}\approx4000$\,K and 0.1\,dex at $T_{\mathrm{eff}}\approx7000$\,K; we apply their calibration to our dwarf training sample.

For M dwarfs, we adopt stellar parameters from \cite{2024MNRAS.52711866Q}, who calibrated M-dwarf metallicities using wide binaries with F/G/K-type primaries and M-type secondaries. Compared with APOGEE and other F/G/K+M wide-binary calibrations, their measurements show a scatter of $\sim$0.18--0.20\,dex.

\subsection{2MASS} \label{subsec:2MASS}

The Two Micron All Sky Survey (2MASS; \citealt{2006AJ....131.1163S}) provides all-sky near-infrared imaging in the $J$ (1.25\,$\mu$m), $H$ (1.65\,$\mu$m), and $K_s$ (2.16\,$\mu$m) bands. The 2MASS Point Source Catalog includes photometry and astrometry for 471 million sources, with 10\,$\sigma$ point-source depths of $J=15.8$, $H=15.1$, and $K_s=14.3$\,mag.

\subsection{WISE}\label{subsec:WISE}

The Wide-field Infrared Survey Explorer (WISE; \citealt{2010AJ....140.1868W}) is an all-sky mid-infrared survey with four bands ($W1$--$W4$) centered at 3.4, 4.6, 12, and 22\,$\mu$m. The WISE Source Catalog contains over 563 million pointlike and resolved sources. Owing to the lower sensitivity of $W3$ and $W4$, we use only $W1$ and $W2$ photometry.

\subsection{miniJPAS Synthetic Magnitudes}\label{subsec:miniJPAS}

The miniJPAS survey was conducted at the Observatorio Astrof\'isico de Javalambre (OAJ; \citealt{2014SPIE.9149E..1IC}) using the JPAS-Pathfinder (JPF) camera on the JST/T250 telescope. JPF employs a 60-filter system comprising 54 narrow-band filters (3785--9100\,\AA; FWHM $\approx145$\,\AA; spacing $\Delta\lambda\approx100$\,\AA), two specialized broadband filters ($J0348/uJAVA$ and $J1007$), and four SDSS-like broadband filters ($uJPAS$, $g$, $r$, $i$). Optimized for low-resolution spectroscopy ($R\sim60$), this system resolves key stellar absorption features (\citealt{2012SPIE.8450E..3SM}).

Following \citet{2023ApJS..269...58X}, we convolve the corrected Gaia XP spectra (and their flux uncertainties) with the miniJPAS filter transmission curves to compute synthetic AB magnitudes (\citealt{1983ApJ...266..713O}; \citealt{1996AJ....111.1748F}) and associated uncertainties in all 60 bands. The transmission curves are available via the Spanish Virtual Observatory Filter Profile Service\footnote{\url{http://svo2.cab.inta-csic.es/theory/fps/index.php?mode=browse&gname=OAJ&gname2=JPAS}}. 
These synthetic miniJPAS magnitudes were generated specifically for this analysis and are not released as a standalone data product with this paper; however, Gaia XP--based synthetic photometry in the miniJPAS passbands is publicly available via the BEst STars (BEST) database\footnote{\url{https://nadc.china-vo.org/data/best/}} \citep{2026NatSD..13..265X}.

\subsection{Extinction}\label{subsec:Extinction}

All subsequent colors and magnitudes are dereddened, intrinsic values using the following reddening prescriptions:
\begin{enumerate}[label=\arabic*., noitemsep]
    \item For the training sample, and for models where extinction is held fixed (Section~\ref{subsec:Model Construction}), we adopt $E(B-V)$ from the \citet{1998ApJ...500..525S} (SFD98) dust map. We apply the constant extinction coefficients of \citet{2023ApJS..264...14Z} for Gaia ($G$, $G_{\mathrm{BP}}$, $G_{\mathrm{RP}}$), 2MASS, and WISE photometry, and use the \citet{2016ApJ...821...78S}-based miniJPAS reddening coefficients\footnote{Stored in the miniJPAS.Filter table: \url{https://archive.cefca.es/catalogues/minijpas-pdr201912/vo\_services.html}} for the synthetic magnitudes.
    \item For models that fit extinction as a free parameter (Section~\ref{subsec:Model Construction}), we use the extinction coefficients from \citet{2024ApJ...971..127Z} (Appendix~\ref{app:extinction-coeffs}) and infer extinction directly from the model fit, independent of SFD98.
\end{enumerate}

\subsection{Training Sample Selection}\label{subsec:training_selection}

\begin{figure}
    \centering
    \includegraphics[width=1\linewidth]{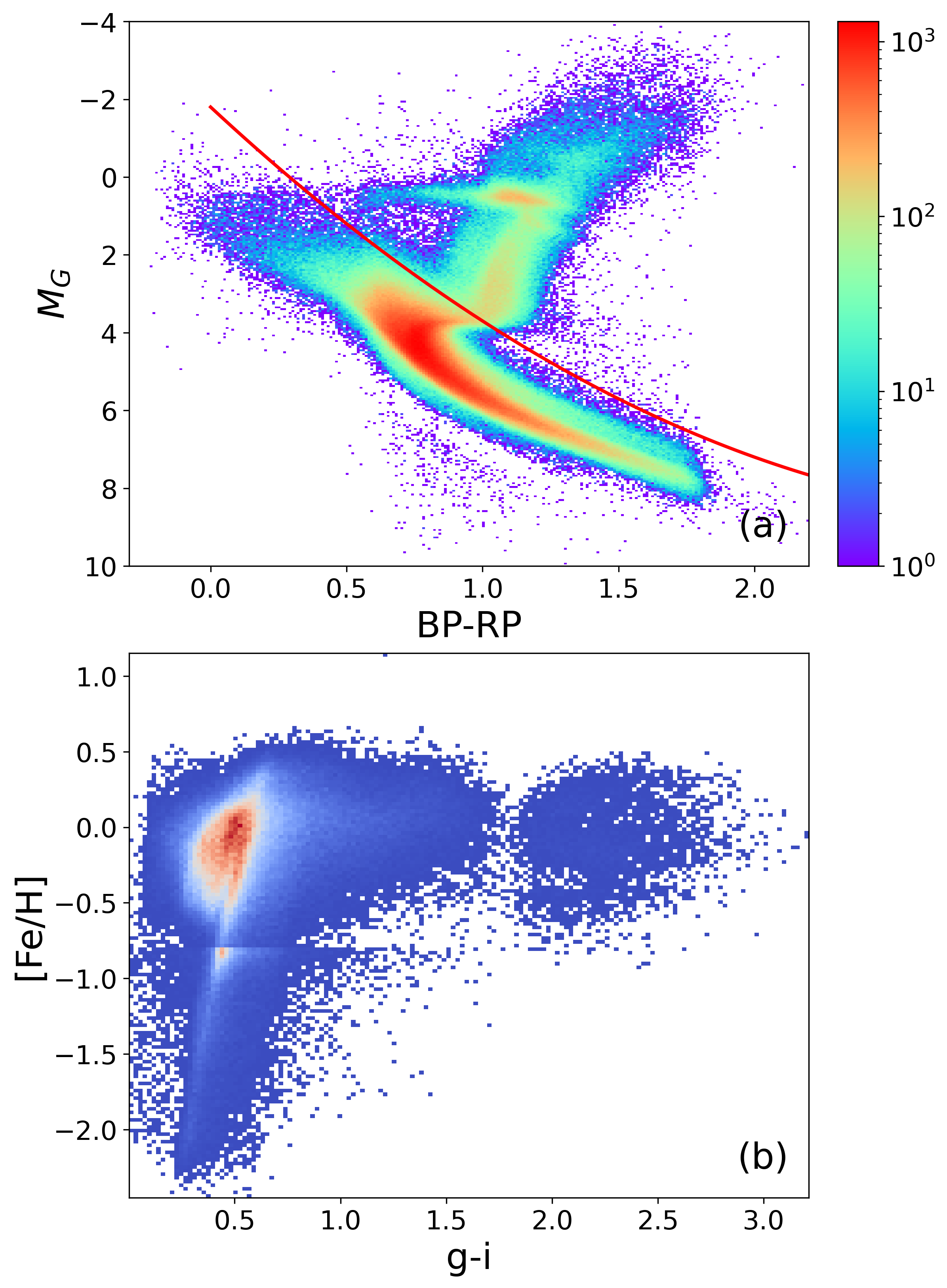}
    \caption{Panel (a) shows the H--R diagram of the LAMOST AFGK sample, color-coded by source density. The red curve separates dwarfs and giants following \citet{2022ApJS..258...44X}, $M_G=-(BP-RP)^2+6.5(BP-RP)-1.8$. Panel (b) shows the distribution of the selected AFGK and M-dwarf training samples in the $\mathrm{[Fe/H]}$--$(g-i)$ plane after applying the selection criteria.}
    \label{fig:hrd_and_gi_feh}
\end{figure}

For sources with multiple LAMOST observations, we retain only the highest signal-to-noise spectrum. We focus on dwarf stars and apply empirical cuts from \citet{2022ApJS..258...44X} to select dwarfs (panel~(a) of Figure~\ref{fig:hrd_and_gi_feh}). We then construct high-quality training samples for AFGK and M dwarfs using a set of criteria applied to all candidate dwarfs, together with additional sample-specific cuts to reflect the different parameter ranges and data-quality characteristics of the two populations.

\noindent\textbf{Common criteria for all dwarf stars.}
\begin{enumerate}
    \item $|b|>20\degree$ and $E(B-V)_{\mathrm{SFD98}}<0.05$: restrict to high-latitude, low-extinction sources to minimize reddening.
    \item $\mathrm{RUWE}<1.2$: exclude sources with elevated Renormalized Unit Weight Error \citep{2021A&A...649A...4L}, which is often associated with unresolved binaries \citep{2020MNRAS.496.1922B, 2021ApJ...907L..33S}.
    \item Gaia XP correction validity: require the first two entries of \texttt{Caution} to be 0, ensuring sources fall within the regime where the correction is applicable \citep{2024ApJS..271...13H}.
    \item WISE quality: require \texttt{cc\_flags[0:2]}$=\texttt{00}$ and \texttt{ext\_flags}$=0$ to exclude artifacts and extended sources and ensure clean $W1/W2$ photometry.
\end{enumerate}

\noindent\textbf{Criteria for AFGK stars.}
\begin{enumerate}
    \item LAMOST SNR$_g>20$: ensure reliable spectroscopic parameters.
    \item $4000<T_{\mathrm{eff}}<7000$\,K: restrict to the AFGK dwarf regime.
    \item Gaia quality: require \texttt{phot\_bp\_rp\_excess\_factor}$<0.09\,(BP_{\mathrm{XPSP}}-RP_{\mathrm{XPSP}})+1.15$. This quantity, $C=(I_{BP}+I_{RP})/I_G$, flags background and contamination in the $BP/RP$ photometry and spectra. 
    In crowded fields, flux from nearby sources can contaminate the BP/RP spectra, shifting a source away from the single-star locus in color--color space. Such contamination can bias the inferred parameters, inflate their uncertainties, and can mimic binarity because the binary-star model can partially absorb the extra flux. We therefore apply this cut to all samples analyzed in this work.
    
    \item Magnitude limits: require $12\leq G\leq14$ for the main sample, and extend to $G\leq16$ for metal-poor dwarfs with $\mathrm{[Fe/H]}\leq-0.8$.
    \item Color cut: require $(g-i)>0$ to exclude bluer sources outside the adopted AFGK range.
\end{enumerate}

\noindent\textbf{Criteria for M stars.}
\begin{enumerate}
    \item LAMOST SNR$_i>20$: ensure reliable parameter estimates.
    \item Gaia quality: require \texttt{phot\_bp\_rp\_excess\_factor}$<0.125\,(BP_{\mathrm{XPSP}}-RP_{\mathrm{XPSP}})+1.135$.
    \item Magnitude limits: require $12\leq G\leq15$ for $\mathrm{[Fe/H]}>-0.4$, and extend to $G\leq16$ for $\mathrm{[Fe/H]}\leq-0.4$.
\end{enumerate}

With these cuts, the training sample comprises 277,813 LAMOST AFGK dwarfs and 4,728 M dwarfs from \citet{2024MNRAS.52711866Q}. Their distributions in the $\mathrm{[Fe/H]}$--$(g-i)$ plane are shown in panel~(b) of Figure~\ref{fig:hrd_and_gi_feh}. 
These stringent cuts aim to calibrate the metallicity-dependent stellar loci and color--absolute-magnitude relations with maximal fidelity, limiting the effects of spectroscopic-parameter uncertainties, interstellar extinction, and photometric contamination on the intrinsic models.

\section{METHODS} \label{sec:METHODS}

\subsection{Metallicity-dependent Stellar Loci} \label{subsec:stellar loci}

Following \citet{2023MNRAS.518.2018Y}, we fit 64 metallicity-dependent stellar loci using the training sample. For each color $X-r$ (where $X$ is any miniJPAS, 2MASS, or WISE band except $r$), we model the locus with a two-dimensional polynomial,
\begin{equation}
X-r = f\bigl((g-i),\,\mathrm{[Fe/H]}\bigr).
\end{equation}

In color--color space, AFGK and M dwarfs follow distinct loci with different metallicity sensitivities, so we adopt a two-stage fit. 
Stage~I spans $0<(g-i)<1.85$ and Stage~II spans $1.45<(g-i)<3.0$; in both stages, we reject outliers via iterative $3\sigma$ clipping.

For inference, we use Stage~I for $(g-i)<1.6$ and Stage~II for $(g-i)>1.7$. In the transition region $1.6\leq(g-i)\leq1.7$, we linearly blend the two fits with weights
\begin{equation}
w_1 = 1 - \frac{(g-i)-1.6}{0.1},\qquad w_2 = 1-w_1,
\end{equation}
which ensures continuity across the boundary.

We select the polynomial order separately for each color. Because the bluer Stage~I colors are more metallicity-sensitive, we adopt higher-order polynomials where justified, and lower-order polynomials for colors with weaker metallicity dependence to mitigate overfitting.

The M-dwarf training sample is limited to $\mathrm{[Fe/H]}\geq-1$, so the Stage~II loci are not reliable below this value. We therefore impose $\mathrm{[Fe/H]}=-1$ as the lower metallicity boundary for Stage~II in subsequent analyses.

The fitted loci are shown in Appendix~\ref{app:stellar-locus-coeffs} (Figure~\ref{fig:loci}). Inspection of the fit residuals reveals no evident systematic dependence on either color or metallicity over the fitted ranges. Stage~I coefficients are listed in Tables~\ref{tab:table_2052_1}--\ref{tab:table_2052_2} and \ref{tab:table_2063_all_new}, and Stage~II coefficients are provided in Table~\ref{tab:table_righe}.

\subsection{Metallicity-Dependent Color--Absolute-Magnitude Relation} \label{subsec:Metallicity-Dependent Color-Absolute Magnitude Relation}

To better constrain binary parameters at high mass ratios using HRD positions, our model requires an accurate color--absolute-magnitude relation for single main-sequence stars.
Using photogeometric distances ($\mathrm{rpgeo}$) from \citet{2021AJ....161..147B}---which combine Gaia parallaxes, colors, and magnitudes in a Bayesian framework---we computed absolute $r$-band magnitudes, $M_r$.
We binned the training sample by metallicity and, within each bin, retained only stars below an empirical boundary before fitting a two-dimensional polynomial to the remaining sample. The selection is illustrated in Appendix~\ref{app:single-star-selection} for AFGK and M dwarfs (Figures~\ref{fig:cutb_afgk} and \ref{fig:cutb_md}, respectively).

To improve the fit quality, we use a piecewise model with a breakpoint at \((g-i)=1.42\) and fit the data separately on either side, rejecting outliers in each segment via iterative $3\sigma$ clipping. Within the transition region, \(1.37 < (g-i) < 1.47\), we linearly blend the two fits in the same way as for the stellar loci (Section~\ref{subsec:stellar loci}) to ensure continuity. The resulting relation is given by

\begin{equation}
M_r(x,y) = \left\{
    \begin{aligned}
        & \begin{aligned}
            1.6127x^4 -0.0888x^3y +0.3036x^2y^2 &\\
            -5.2232x^3 +0.9291x^2y -0.5176xy^2 &\\
            +4.3751x^2 -1.4171xy +0.0825y^2 &\\
            +2.7092x -0.1811y +2.6236 & \\
            for~~ x < 1.42 
            \end{aligned} \\[2ex]
        & \begin{aligned}
            0.3534x^3 +0.0470x^2y +0.0392xy^2 &\\
            -1.1119x^2 -0.3185xy -0.1541y^2 &\\
            +2.8759x -0.2624y +4.0678 & \\
            for ~~ x \geq 1.42
            \end{aligned}
    \end{aligned}
\right.
\label{eq:absolute_magnitude}
\end{equation}

\begin{figure}
    \centering
    \includegraphics[width=1\linewidth]{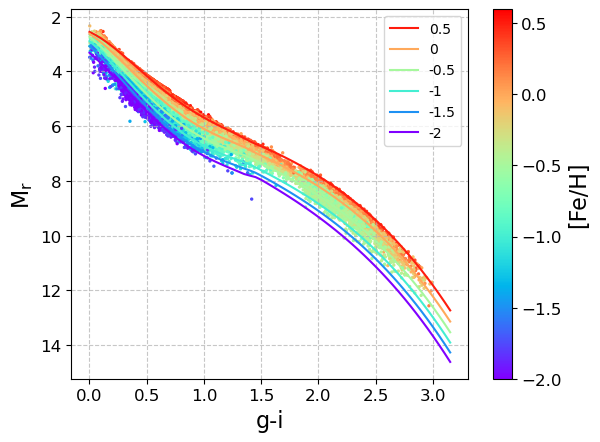}
    \caption{HRD of the training sample after removing binaries and main-sequence turnoff stars. The colored curves show the color--absolute-magnitude relation at different \(\mathrm{[Fe/H]}\) values (as indicated in the legend).}
    \label{fig:color_mag_rel}
\end{figure}

Here, \(x\) denotes the \((g-i)\) color and \(y\) denotes \(\text{[Fe/H]}\). The fitted relation has an intrinsic scatter of 0.127~mag, which we adopt as the uncertainty floor for the HRD constraint in the subsequent model fitting. The fit is shown in Figure~\ref{fig:color_mag_rel}.

\subsection{Model Construction} \label{subsec:Model Construction}

In this subsection, we describe how we construct forward models for single stars and unresolved binaries.

For single stars, the intrinsic photometric SED is parameterized by \(\mathrm{[Fe/H]}\), the apparent \(r\)-band magnitude, and the intrinsic \((g-i)\) color. Given \((g-i)\) and \(\mathrm{[Fe/H]}\), we predict the 64 colors from the stellar loci; together with the \(r\)-band magnitude, these define the 65-band photometric SED (hereafter SED).

For unresolved binaries, the intrinsic SED is parameterized by a common \(\mathrm{[Fe/H]}\), and by \((r_1,(g-i)_1)\) and \((r_2,(g-i)_2)\) for the primary and secondary, respectively. In any band, the combined magnitude is
\[
m_{\mathrm{binary}} = -2.5\log_{10}\left(10^{-0.4m_1}+10^{-0.4m_2}\right),
\]
where \(m_1\) and \(m_2\) are the component magnitudes predicted by the corresponding single-star loci models.

\begin{figure}
    \centering
    \includegraphics[width=1\linewidth]{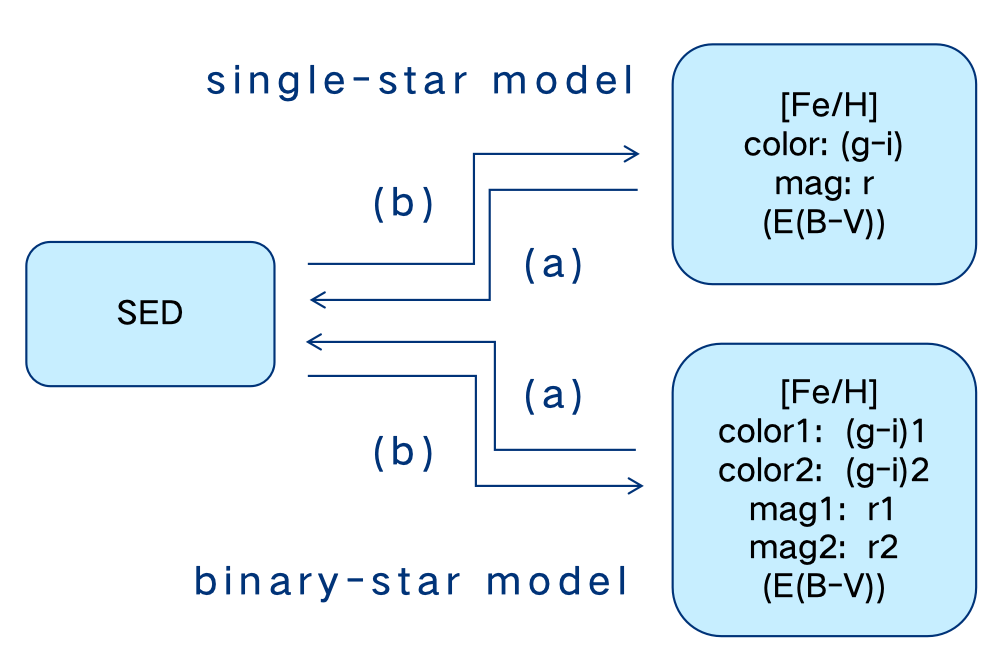}
    \caption{Schematic overview of the workflow: (a) forward model that generates an SED from input parameters; (b) inverse model that constrains parameters by fitting the observed SED with MCMC.}
    \label{fig:model}
\end{figure}

We optionally include \(E(B-V)\) as a free parameter in both the single-star and binary-star models. When included, we apply extinction to the model SEDs using band-specific reddening coefficients \(k_x\) (all 65 bands) derived from the extinction curve of \citet{2024ApJ...971..127Z} (Table~\ref{tab:kx_ebv015_teff5500_logg45}), enabling the models to represent reddened sources. Figure~\ref{fig:model} provides a schematic overview of the full modeling framework.

\begin{figure*}
    \centering
    \includegraphics[width=\linewidth,height=0.96\textheight,keepaspectratio]{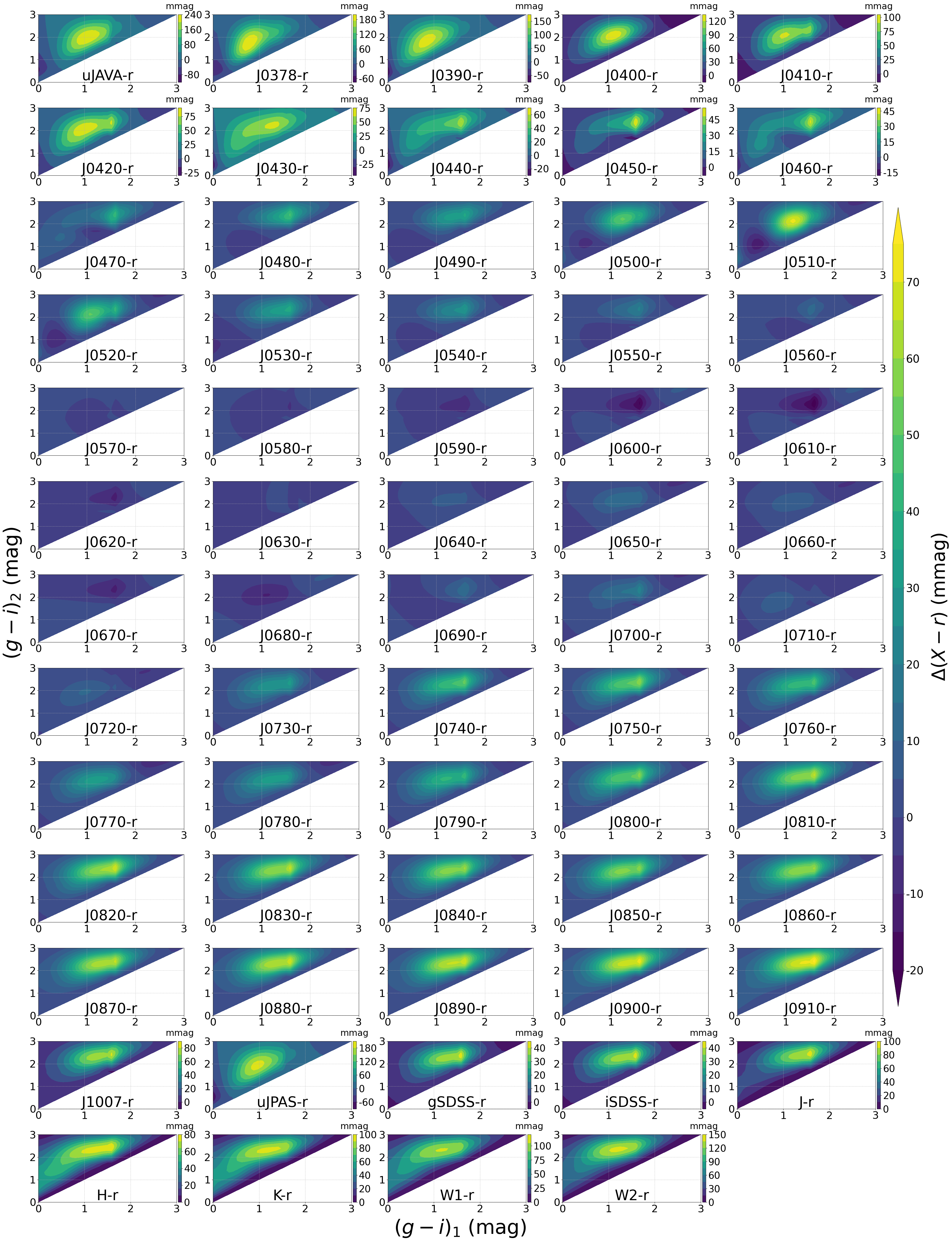}
    \caption{Differences between the colors predicted by the metallicity-dependent stellar loci at the system composite $(g-i)$ and the composite binary colors, $\Delta(X-r)\equiv(X-r)_{\mathrm{predicted}}-(X-r)_{\mathrm{obs}}$, for the color $X-r$ labeled in each panel. The residuals are shown as a function of the primary and secondary $(g-i)$ colors for main-sequence binaries with $\mathrm{[Fe/H]}=0$. The central panels share a common color bar; the surrounding panels have individual color bars. All color bars are in mmag.}
    \label{fig:differences_contour}
\end{figure*}

To motivate multi-band SED fitting and the HRD constraint, we quantify the sensitivity of different color loci to unresolved binarity. Under idealized conditions (zero photometric uncertainty, no extinction, and fixed $\mathrm{[Fe/H]}=0$), we compute residuals between composite binary colors and single-star locus predictions. We define the residual in a given locus as
\[
\Delta (X-r) = (X-r)_{\mathrm{predicted}} - (X-r)_{\mathrm{obs}}.
\]
Here, $(X-r)_{\mathrm{obs}}$ is the true composite color of a binary system from our binary forward model, evaluated on a grid in component colors $(g-i)_1$ and $(g-i)_2$. The predicted color $(X-r)_{\mathrm{predicted}}$ is obtained from the single-star locus using the system's composite $(g-i)_{\mathrm{obs}}$ and the fixed $\mathrm{[Fe/H]}=0$. We map $\Delta(X-r)$ as a function of the component colors in Figure~\ref{fig:differences_contour}.

We find that unresolved binaries systematically deviate from the single-star loci, with the magnitude of the deviation depending on both the chosen color locus and the primary/secondary color combination. Infrared loci show a larger dynamic range in $\Delta(X-r)$ than optical loci, implying greater sensitivity to unresolved binarity; accordingly, including infrared bands should improve our ability to probe binaries over a wider range of mass ratios.

The forward models described above map physical parameters to a predicted SED. To infer these parameters from the data, for each observed SED we fit both the single-star and unresolved-binary model families using Markov chain Monte Carlo (MCMC; \citealt{2013PASP..125..306F}). Assuming Gaussian uncertainties, we quantify the agreement between the observed and model photometry, together with the HRD constraint, with the statistic
\begin{equation}
\chi^2 = 3\sum_{j=1}^{65} \frac{\left(m_j^{\mathrm{obs}}-m_j^{\mathrm{mod}}\right)^2}{\sigma_j^2}
+ \frac{\left(M_r^{\mathrm{obs}}-M_r^{\mathrm{mod}}\right)^2}{\sigma_{M_r^{\mathrm{obs}}}^2+0.127^2}.
\end{equation}
Here, $m_j^{\mathrm{obs}}$ and $m_j^{\mathrm{mod}}$ are the observed and model magnitudes in band $j$, and $\sigma_j$ is the corresponding photometric uncertainty. The quantity $M_r^{\mathrm{obs}}$ is the observed absolute $r$-band magnitude derived from the apparent magnitude and distance estimate, and $M_r^{\mathrm{mod}}$ is the model absolute magnitude. For the single-star model,
\begin{equation}
M_{r,\mathrm{s}}^{\mathrm{mod}} = M_r\!\left((g-i)_{\mathrm{s}},[\mathrm{Fe/H}]_{\mathrm{s}}\right),
\end{equation}
whereas for the binary-star model,
\begin{equation}
M_{r,\mathrm{b}}^{\mathrm{mod}} = -2.5\log_{10}\!\left(10^{-0.4 M_{r,1}^{\mathrm{imp}}} + 10^{-0.4 M_{r,2}^{\mathrm{MS}}}\right).
\end{equation}

The uncertainty in the observed absolute magnitude is
\begin{equation}
\sigma_{M_r^{\mathrm{obs}}} = \sqrt{\sigma_r^2 + \left(\frac{5}{\ln 10}\frac{\sigma_d}{d}\right)^2},
\end{equation}
where $\sigma_r$ is the uncertainty in the apparent $r$-band magnitude, $d$ is the distance, and $\sigma_d$ is its uncertainty.
We set $d$ to the median of the photogeometric distance posterior from \citet{2021AJ....161..147B}, and define $\sigma_d=\sqrt{\left[(d_{84}-d)^2+(d-d_{16})^2\right]/2}$, where $d_{16}$ and $d_{84}$ are the 16th and 84th percentiles. 
The 0.127 mag term in the denominator of the $M_r$ constraint represents the intrinsic scatter of the empirical color--absolute-magnitude relation.

We multiply the 65-band photometric term by a factor of $3$ so that the SED information is not underweighted relative to the HRD constraint; this choice is motivated by the possibility that the reported photometric uncertainties are mildly conservative.

For the single-star model, we apply a modified HRD constraint for blue stars with $(g-i)<0.75$, where the main-sequence turn-off (MSTO) broadens the locus. Defining
\begin{equation}
\Delta M_r \equiv M_r^{\mathrm{obs}} - M_r^{\mathrm{mod}},
\end{equation}
we use a one-sided penalty in the absolute-magnitude term to avoid over-penalizing stars that lie above the empirical main sequence:
\begin{equation}
(\Delta M_r)^2 \;\rightarrow\;
\begin{cases}
0, & \Delta M_r > -1.8,\\
(\Delta M_r+1.8)^2, & \Delta M_r < -1.8.
\end{cases}
\end{equation}
This choice avoids penalizing MSTO stars for lying above the empirical main sequence in the single-star fits.

We sample the parameters with MCMC, adopting the likelihood $\mathcal{L}\propto\exp(-\chi^2/2)$. To ensure physically plausible binary solutions, we impose the following priors.
\begin{enumerate}
\item \textit{Parameter bounds.} The free parameters are restricted to
\begin{align}
0 < r_1 < 30,\qquad &0 < r_2 < 30,\\
-3 < \mathrm{[Fe/H]} < 1,\\
-0.5 < (g-i)_1 < 3.5,\qquad &-0.5 < (g-i)_2 < 3.5.
\end{align}

\item \textit{Component ordering.} The primary is required to be no redder and no fainter than the secondary, within observational uncertainties:
\begin{align}
(g-i)_1 &\le (g-i)_2 + \sigma_{g-i},\\
r_1 &\le r_2 + \sigma_r
\end{align}

\item \textit{Main-sequence consistency and distance modulus.} We assume the secondary lies on the empirical main-sequence color--absolute-magnitude relation, which sets the system distance modulus,
\begin{equation}
\mu = r_2 - M_{r,2}^{\mathrm{MS}}.
\end{equation}
For the primary we define the main-sequence reference magnitude
\begin{equation}
M_{r,1}^{\mathrm{MS}} = M_r\!\left((g-i)_1,\mathrm{[Fe/H]}\right).
\end{equation}
For redder primaries ($ (g-i)_1 \ge 0.75 $), we require the primary to remain close to the main sequence,
\begin{equation}
r_1 - M_{r,1}^{\mathrm{MS}} = \mu.
\end{equation}
For bluer primaries ($ (g-i)_1 < 0.75 $), we allow evolution toward the MSTO (i.e., brighter-than-MS) but enforce
\begin{align}
r_1 &\le M_{r,1}^{\mathrm{MS}}+\mu+0.127,\\
r_1 &\ge M_{r,\mathrm{obs}}+\mu-\sigma_{M_r^{\mathrm{obs}}}-0.127,
\end{align}
which prevents the primary from exceeding the total system brightness.
\end{enumerate}

For each source, we run the single-star and binary-star fits independently with MCMC (30 walkers per ensemble; 10{,}000 steps), using a weighted mixture of differential-evolution and snooker proposals (\texttt{DEMove} and \texttt{DESnookerMove}, with weights 0.8 and 0.2; \citealt{2006StCom..16..239T,2008StCom..18..435T}). For the single-star model, we run one ensemble per source, initialized at $\mathrm{[Fe/H]}=-0.2$, $(g-i)=(g-i)_{\mathrm{sys}}$, and $r=r_{\mathrm{obs}}$. Because the secondary parameters can be multimodal, a single binary-model ensemble may become trapped in a local $\chi^2$ minimum; we therefore run 2--5 independent binary ensembles per source that differ only in the initial $(g-i)_2$ (Table~\ref{tab:mcmc_init}). The remaining binary parameters are initialized at $\mathrm{[Fe/H]}=-0.5$, $(g-i)_1=(g-i)_{\mathrm{sys}}$, and $r_1=r_{\mathrm{obs}}+0.75$, with $r_2$ set by enforcing a common distance modulus. For each ensemble (both models), we estimate the integrated autocorrelation time $\tau$ with \texttt{emcee} and discard the first $5\tau_{\max}$ steps as burn-in, where $\tau_{\max}$ is the maximum $\tau$ across parameters. For the binary model, we compute $\chi^2$ at the posterior median of each ensemble and adopt the ensemble with the lowest $\chi^2$. For the adopted binary ensembles, the typical (median) $\tau_{\max}$ is $\approx 280$ steps across all samples analysed in this paper (278--286 for the simulated, LAMOST, M67, and Gaia samples), driven primarily by the secondary parameters $(g-i)_2$ and $r_2$; the single-star model mixes much faster, with a median $\tau_{\max}$ of 16--20 steps. As an additional check, we split each retained chain into two halves and compare the resulting posterior medians. For $>99.8\%$ of sources in every sample, the medians of all parameters (including the secondary) agree within $1\sigma$, where $\sigma$ is half of the 16th--84th percentile range of the full retained chain. For multimodal posteriors, the quoted uncertainties describe only the adopted mode. 
We summarize each parameter by the median of its marginal posterior and quote the 16th and 84th percentiles as a central 68\% credible interval.

We then compare the two model families using their minimum $\chi^2$ values and adopt the lower-$\chi^2$ fit as the preferred interpretation of the observed SED. We take the parameters from the preferred model as the final source parameters, and derive stellar masses (and thus binary mass ratios) together with their asymmetric uncertainties using the PARSEC evolutionary models (Appendix~\ref{app:mass-determination}).

\begin{deluxetable}{lll}
\tabletypesize{\footnotesize}
\tablecaption{Initial $(g-i)_2$ Values of the Independent MCMC Ensembles for the Binary Model\label{tab:mcmc_init}}
\tablewidth{0pt}
\tablehead{
\colhead{$(g-i)_{\mathrm{sys}}$} & \colhead{Initial $(g-i)_2$} & \colhead{$N_{\mathrm{ens}}$}
}
\startdata
$< 0.8$      & $(g-i)_{\mathrm{sys}} + \{0.07, 0.65, 1.35, 2.2\}$, 3.3 & 5 \\
$0.8$--$1.7$ & $(g-i)_{\mathrm{sys}} + \{0.07, 0.65, 1.35\}$, 3.3      & 4 \\
$1.7$--$2.5$ & $(g-i)_{\mathrm{sys}} + \{0.07, 0.6\}$, 3.3             & 3 \\
$\geq 2.5$   & $(g-i)_{\mathrm{sys}} + 0.07$, 3.3                       & 2 \\
\enddata
\end{deluxetable}

\section{RESULTS AND DISCUSSION} \label{sec:RESULTS AND DISCUSSION}

\subsection{Test by Simulated Data} \label{Test by Simulated Data}

In this subsection, we assess the performance of our method using simulated data. We generate a set of mock unresolved binaries with the forward model described in Section~\ref{subsec:Model Construction}. All systems are assigned $\mathrm{[Fe/H]}=-0.1$ and placed at a fixed distance of 500\,pc. Primary colors $(g-i)_1$ are drawn uniformly from $0\le (g-i)_1\le 1.75$, and secondary colors $(g-i)_2$ from $0\le (g-i)_2\le 3$, with $(g-i)_1\le (g-i)_2$. Component absolute magnitudes are then set using the color--absolute-magnitude relation in Section~\ref{subsec:Metallicity-Dependent Color-Absolute Magnitude Relation}. We do not apply interstellar extinction.

To mimic observational uncertainties, we construct empirical error models from the data: we fit a polynomial to the photometric uncertainty as a function of magnitude, and a second polynomial to the distance uncertainty as a function of magnitude and distance. We use these fits to assign uncertainties to each mock system and add Gaussian noise to the photometry and distances.

\begin{figure*}
    \centering
    \includegraphics[width=1\linewidth]{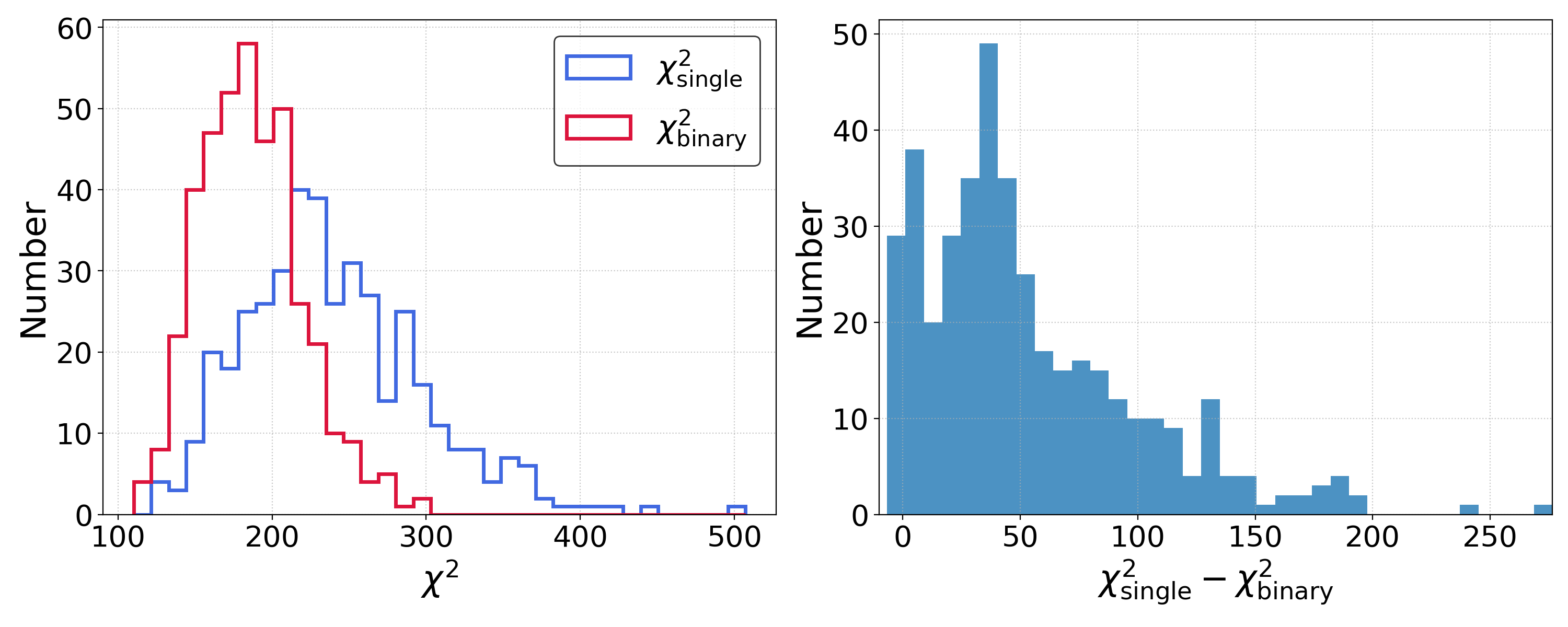}
    \caption{Left: $\chi^2$ distributions for single-star and binary-star fits to the simulated data. Right: distribution of $\Delta\chi^2 \equiv \chi^2_{\rm single}-\chi^2_{\rm binary}$ for the simulated data.}
    \label{fig:simu_chisq}
\end{figure*}

We then fit the simulated sources with both the single-star and binary-star models, holding extinction fixed (i.e., not including it as a free parameter). Figure~\ref{fig:simu_chisq} shows the resulting distributions of $\chi^2$ and $\Delta\chi^2$. Because all simulated sources are unresolved binaries, $\Delta\chi^2$ is predominantly $>0$, indicating that the binary-star model is preferred in nearly all cases.

\begin{figure*}
    \centering
    \includegraphics[width=1\linewidth]{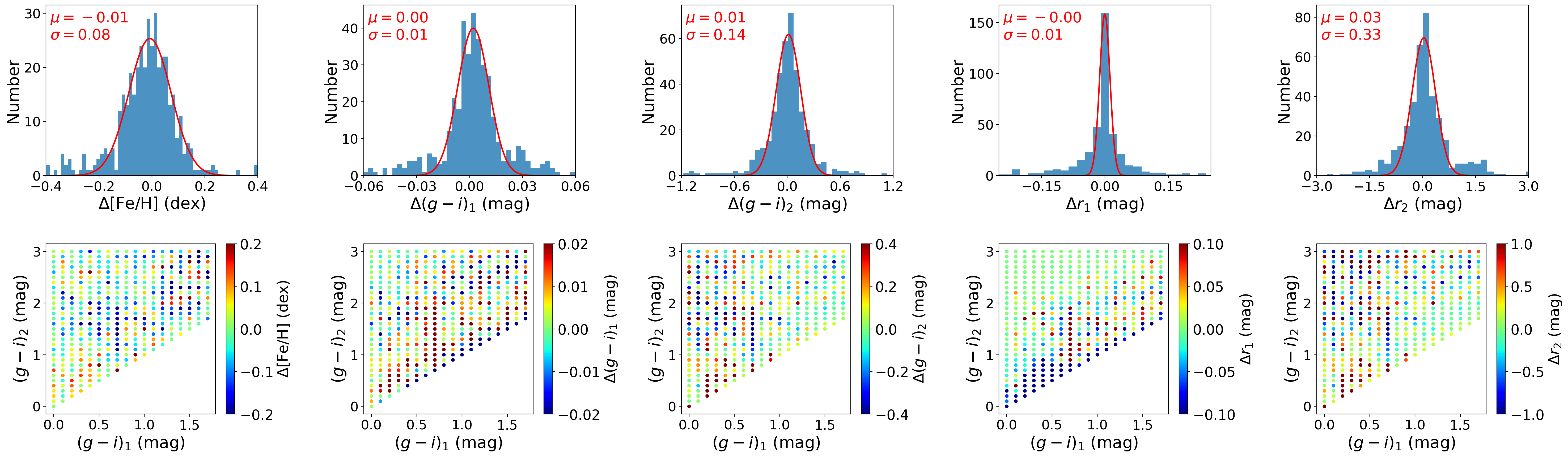}
    \caption{Binary-model fit residuals for the simulated data. Top row: histograms of parameter residuals (recovered minus input). Bottom row: residual structure across the primary--secondary color--color plane.}
    \label{fig:simu_grad}
\end{figure*}

Figure~\ref{fig:simu_grad} shows residuals between the binary-model posterior medians and the true input parameters. Overall, the recovered values track the inputs across the explored parameter space, with only small systematic offsets in metallicity and primary-star parameters, 
but with markedly larger scatter in the secondary color and magnitude ($\sigma=0.14$ and 0.33\,mag, respectively). 
This indicates that the method can recover intrinsic binary parameters to within the expected uncertainties.

\begin{figure*}
    \centering
    \includegraphics[width=1\linewidth]{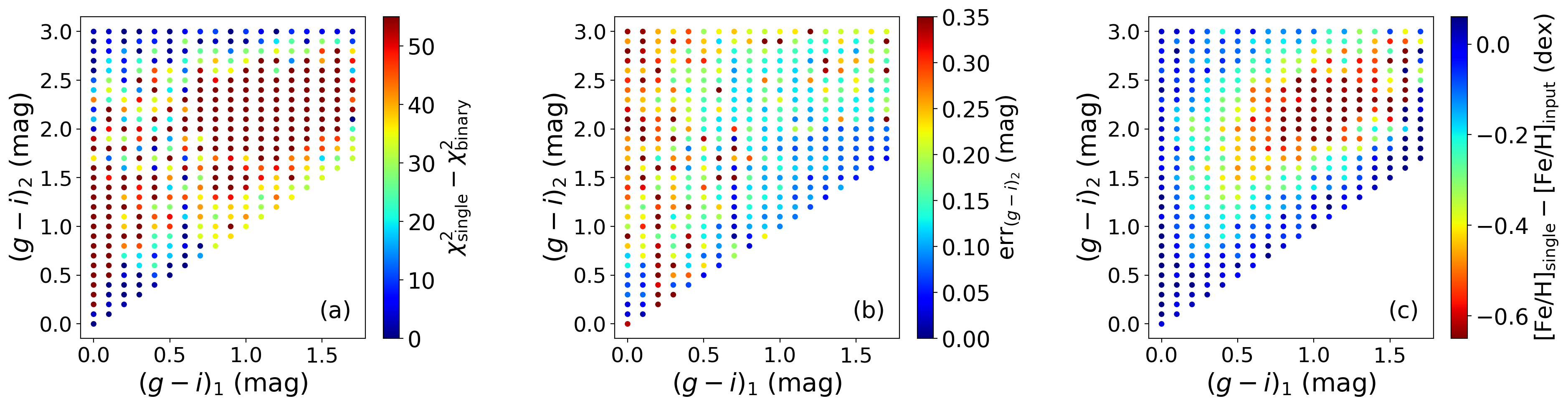}
    \caption{From left to right: $\Delta\chi^2$, secondary-color uncertainty, and metallicity residual $\Delta\mathrm{[Fe/H]}$ across the primary--secondary color--color plane for the simulated data.}
    \label{fig:simu_del_chisq_feh}
\end{figure*}

Figure~\ref{fig:simu_del_chisq_feh}(a) maps $\Delta\chi^2$ across the simulated parameter grid, highlighting where unresolved binaries can be most cleanly separated from single-star fits over a broad range of mass ratios. The high-sensitivity regions align with the color-residual structure in Figure~\ref{fig:differences_contour}, and show that the method performs best for main-sequence binaries with moderate mass ratios. Including the absolute-magnitude term in the likelihood further improves sensitivity to large mass-ratio systems, for which color deviations alone are small.

Figure~\ref{fig:simu_del_chisq_feh}(b) shows the distribution of secondary-color uncertainties. The posterior widths of $(g-i)_2$ increase substantially near the MSTO, consistent with the reduced locus sensitivity in this regime (Figure~\ref{fig:differences_contour}). In addition, the large primary--secondary flux contrast and the rapid structural changes near the MSTO make it difficult to constrain the secondary simultaneously in color and magnitude, leading to larger uncertainties.

Figure~\ref{fig:simu_del_chisq_feh}(c) shows that fitting unresolved binaries with single-star models introduces a systematic metallicity bias: the inferred metallicities are biased low. The companion contaminates the composite SED, displacing the system from the single-star loci in color--color space. As illustrated in Figure~\ref{fig:differences_contour}, at fixed metallicity different combinations of primary and secondary colors can produce similar composite colors, so a binary can fall on an apparently plausible single-star locus yet imply an incorrect metallicity. 

We emphasize that these simulations are intentionally idealized, designed to isolate parameter recoverability under controlled conditions rather than to represent the full diversity of observational environments.

\subsection{Test with LAMOST Data} \label{Test by LAMOST}

\begin{figure}
    \centering
    \includegraphics[width=1\linewidth]{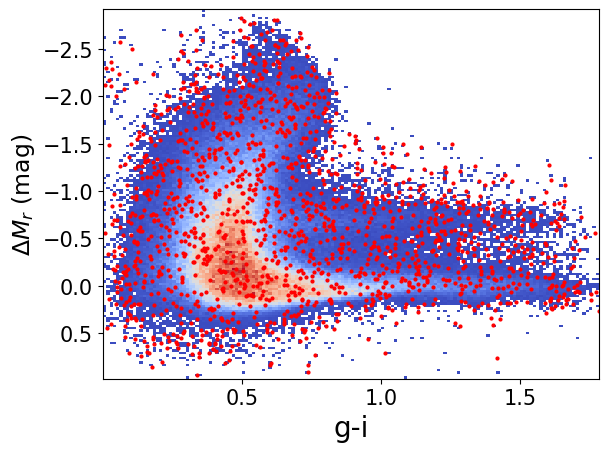}
    \caption{Distribution of the LAMOST test sample on the $\Delta$-HRD. Red points mark the sources selected by sampling in $\Delta$-HRD space, and the background shows the AFGK dwarf training sample.}
    \label{fig:source}
\end{figure}

Here we use a LAMOST DR10 comparison sample to evaluate our fitting framework, focusing on discrimination between the single-star and unresolved-binary hypotheses, the accuracy of photometric metallicity recovery, and the robustness of the extinction estimates.
We select 1{,}600 sources by sampling the residual Hertzsprung--Russell diagram (hereafter $\Delta$-HRD), whose ordinate is defined as $\Delta M_r \equiv M_r^{\mathrm{obs}} - M_r^{\mathrm{mod}}$ (Figure~\ref{fig:source}). Using the $\Delta$-HRD reduces metallicity-driven structure in the HR diagram, making the separation between main-sequence single stars and unresolved main-sequence binaries (which lie above the main sequence) more apparent. The sample follows the same selection criteria as the AFGK dwarf stars in the training set, except that we do not apply a RUWE cut to avoid preferentially excluding unresolved binaries.

\begin{figure*}
    \centering
    \includegraphics[width=1\linewidth]{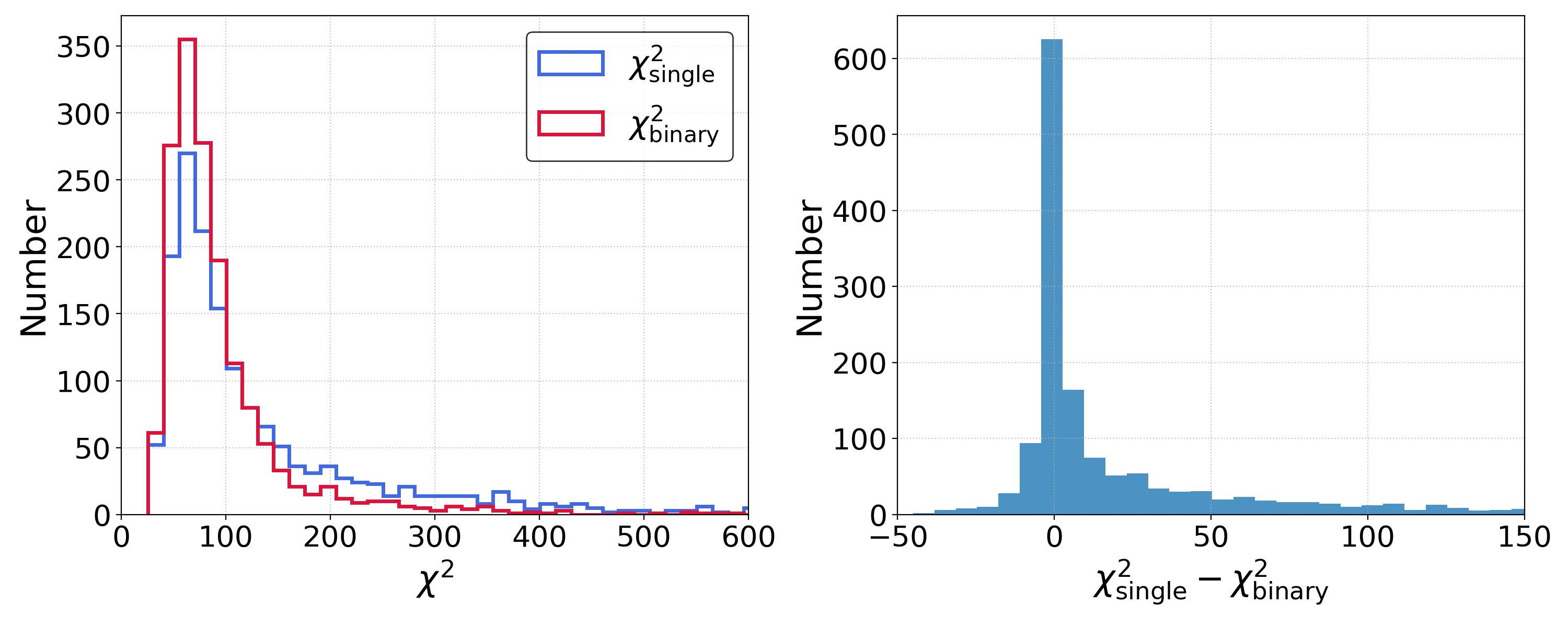}
    \caption{Left: $\chi^2$ distributions for the single-star and binary-star models to the LAMOST test data. Right: distribution of $\Delta\chi^2$ for the LAMOST test data.}
    \label{fig:chi_sq_2model_noebv}
\end{figure*}

We first performed the fits with a model in which $E(B-V)$ was fixed to the SFD98 map (i.e., not treated as a free parameter). Figure~\ref{fig:chi_sq_2model_noebv} shows the resulting distributions of $\chi^2$ and $\Delta\chi^2$ for the single-star and binary-star models. The extended positive tail in $\Delta\chi^2$ is dominated by sources with strong photometric signatures of binarity, whereas objects with $\Delta\chi^2\approx 0$ (or $\Delta\chi^2<0$) are consistent with the single-star hypothesis. This pattern implies that the binary-star model substantially improves the fits when binarity is present, but offers little to no benefit for sources that are already well described by the single-star model.

\begin{figure*}
    \centering
    \includegraphics[width=1\linewidth]{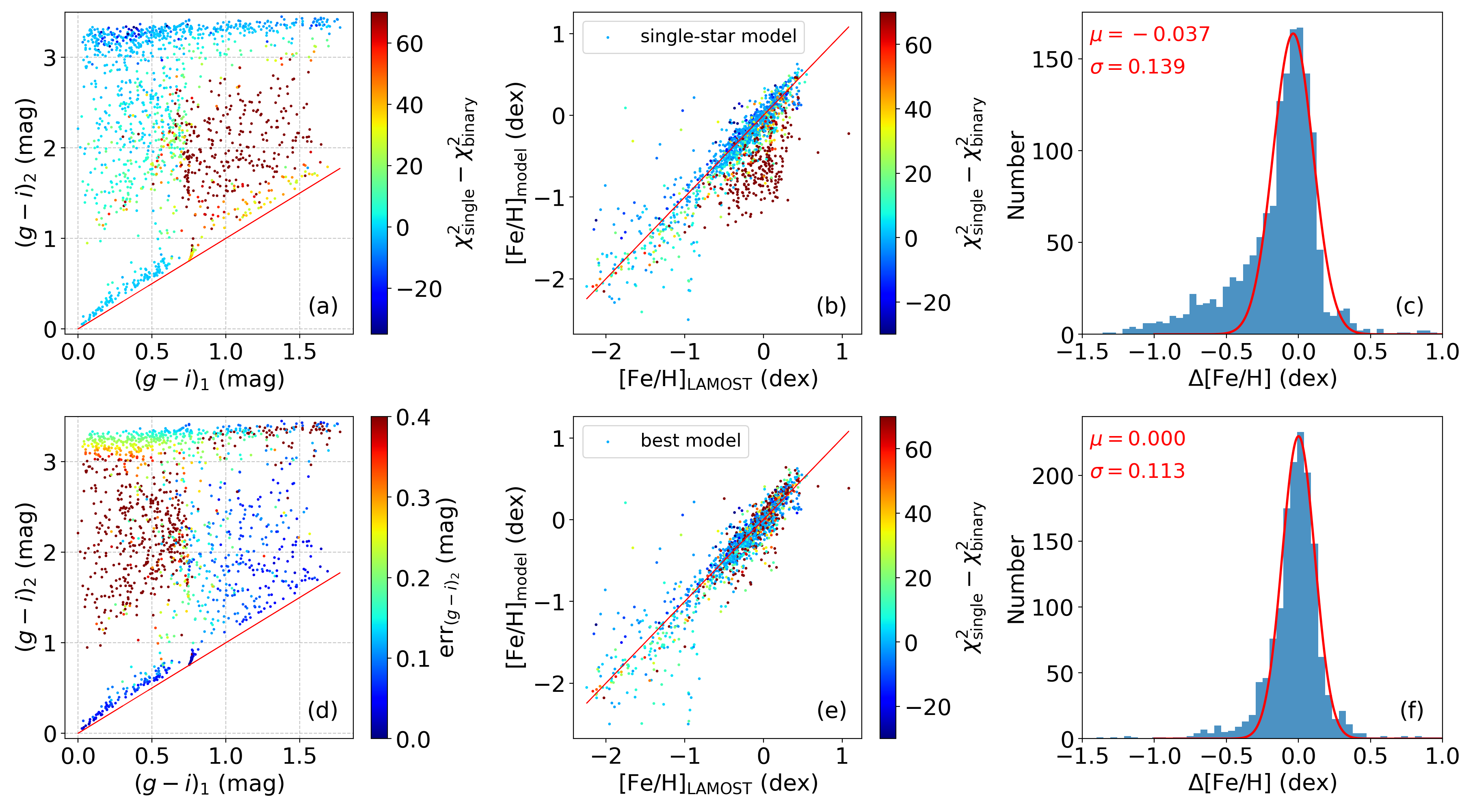}
    \caption{(a) Primary vs. secondary colors $(g-i)$ from the binary-star fits for the LAMOST test data, colored by $\Delta\chi^2$. (b) Single-star photometric metallicities versus LAMOST, colored by $\Delta\chi^2$. (c) Residuals $\Delta\mathrm{[Fe/H]}\equiv \mathrm{[Fe/H]_{model}}-\mathrm{[Fe/H]_{LAMOST}}$ for the single-star model; the red curve shows the Gaussian fit. (d) Same as (a), but colored by secondary-color uncertainty. (e) Best-fit photometric metallicities versus LAMOST (same color scheme as in panel~(b)). (f) Residuals $\Delta\mathrm{[Fe/H]}$ for the best-fit model; red curve: Gaussian fit.}
    \label{fig:feh_vslm_noebv}
\end{figure*}

To guard against catastrophic fits, we discard sources with $\chi^2>200$ in both the single-star and binary-star models. After this quality cut, 1,499 sources remain, and their fitting results are shown in Figure~\ref{fig:feh_vslm_noebv}.

Panel (a) shows the color distribution of the primary and secondary components inferred from the binary-star fits. The color scale encodes $\chi^2_{\mathrm{single}}-\chi^2_{\mathrm{binary}}$, with larger values indicating a stronger preference for the binary-star model. For many sources with $\Delta\chi^2\approx 0$, the binary-star fit converges to a configuration with one luminous component and one very faint companion, producing a pile-up at $(g-i)_2>3$.
Panels (b) and (c), and panels (e) and (f), compare the metallicities from the single-star and best-fit models, respectively, with the LAMOST spectroscopic measurements. In the single-star fits, sources with large $\Delta\chi^2$ (likely unresolved binaries) show a systematic metal-poor bias relative to LAMOST, consistent with the binary-contamination effects reported by \citet{2015ApJ...803...13Y}. In contrast, the best-fit model agrees well with LAMOST: fitting a Gaussian to the residuals $\Delta[\mathrm{Fe/H}]$ yields an offset of $0.00$ and a dispersion of 0.113~dex (panel~f). 
Throughout this paper, unless otherwise stated, the ``offset'' and ``scatter'' of a residual distribution refer to the mean $\mu$ and standard deviation $\sigma$ of the best-fitting Gaussian. 
Overall, these comparisons highlight unresolved binarity as a major source of bias in photometric metallicities and show that explicitly modeling binaries substantially improves agreement with the LAMOST metallicity scale.

\begin{figure*}
    \centering
    \includegraphics[width=1\linewidth]{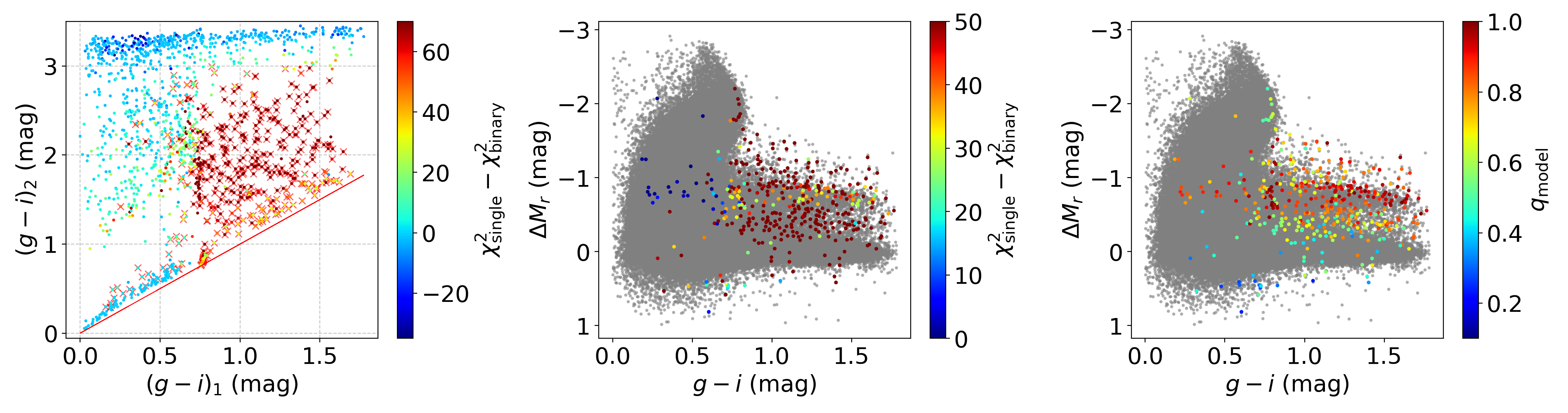}
    \caption{Left: sources selected as candidate binaries (red crosses). Middle: their locations on the $\Delta$-HRD, colored by $\Delta\chi^2$. Right: the same sources, colored by the inferred mass ratio $q$.}
    \label{fig:q_noebv}
\end{figure*}

To assess our ability to recover mass ratios, we select sources that are well fit by the binary-star model and examine how the inferred mass ratios are distributed on the $\Delta$-HRD. We require $\Delta\chi^2>0$, $\mathrm{err}_{(g-i)_2}<0.2$, and $(g-i)_2<3$ to retain only systems with reliable secondary colors.
In Figure~\ref{fig:q_noebv}, red crosses in the first panel mark the primary and secondary colors of the selected sources, while the second and third panels show their $\Delta\chi^2$ and inferred mass ratios on the $\Delta$-HRD. The distribution indicates that our method is most sensitive to main-sequence--main-sequence binaries with $q\gtrsim0.4$. Within this regime, sources move smoothly from the single-star sequence toward the high-$q$ binary sequence as the inferred mass ratio increases, demonstrating that the recovered mass ratios follow the expected photometric loci. 

We repeated the fits with $E(B-V)$ treated as a free parameter in both the single-star and binary-star models. Allowing for extinction does not substantially affect the inferred metallicities: the photometric metallicities from the preferred model remain consistent with the LAMOST spectroscopic values, with a scatter of 0.127~dex. The fitted extinctions also agree well with the SFD map in the low-extinction regime, with a scatter of 0.019~mag. These results indicate that the inferred metallicities and binary classifications are robust to the additional freedom in extinction.

\begin{figure*}
    \centering
    \includegraphics[width=1\linewidth]{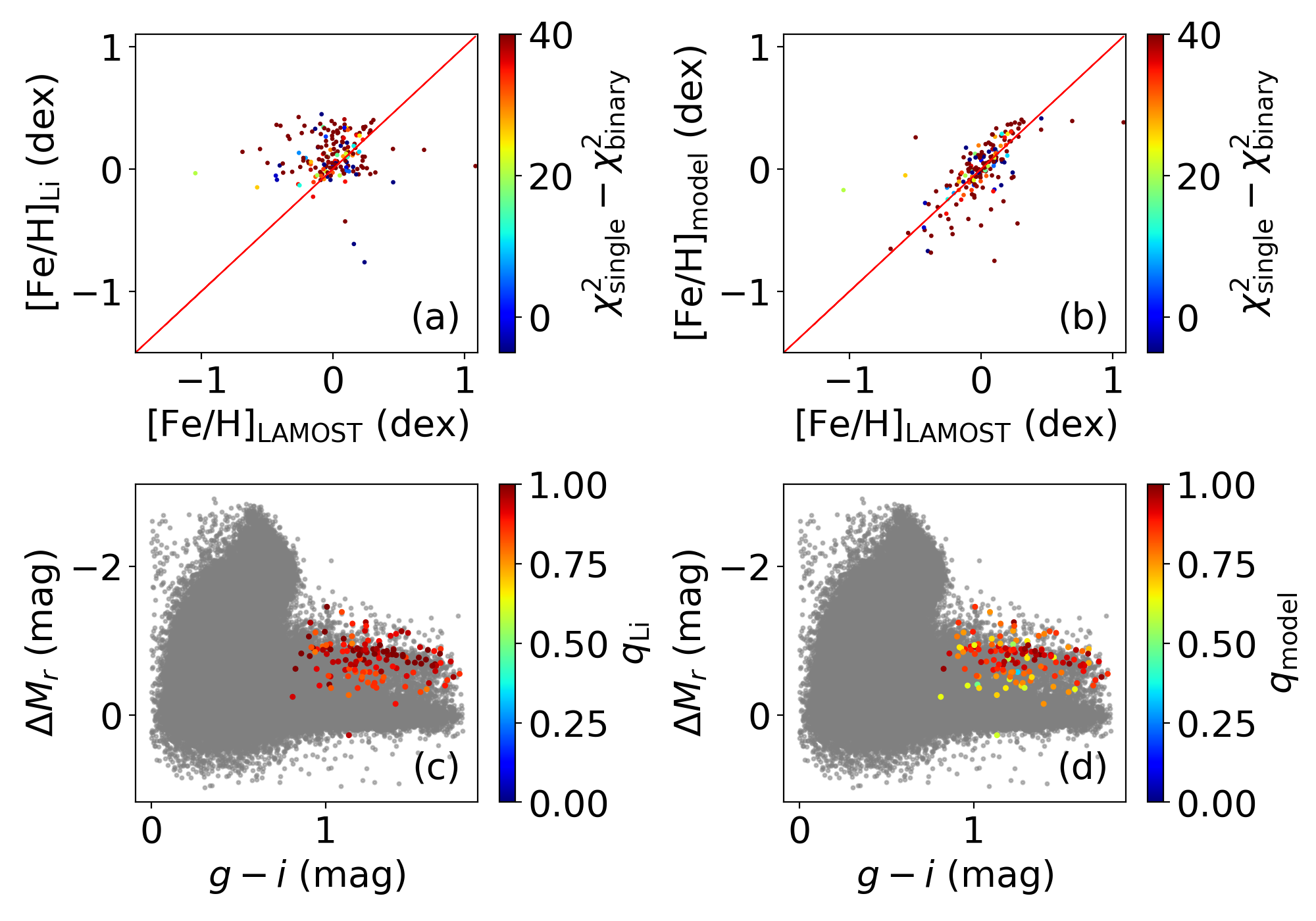}
    \caption{Comparison with the Golden Sample of \citet{2025A&A...704A.126L}. Panels (a) and (b) compare the photometric metallicities from \citet{2025A&A...704A.126L} and from this work, respectively, with the LAMOST spectroscopic metallicities; points are colored by $\Delta\chi^2$. Panels (c) and (d) show the common sources on the $\Delta$-HRD, colored by the mass ratios inferred by \citet{2025A&A...704A.126L} and by this work, respectively. Gray points show the AFGK dwarf training sample.}
    \label{fig:vs_ljd}
\end{figure*}

For an external cross-check, we cross-matched the LAMOST test sample with the Gaia XP--based Golden Sample of \citet{2025A&A...704A.126L} and found 176 sources in common. Of the objects flagged as binaries in that catalog, 159 also satisfy $\Delta\chi^2>0$ in our analysis, indicating broad agreement between the classifications. We then compared the Golden Sample metallicities and mass-ratio estimates with our results and with the LAMOST spectroscopic measurements. Figures~\ref{fig:vs_ljd} (a) and (b) show that our photometric metallicities track the LAMOST values more closely, while panels (c) and (d) indicate that, in the intermediate-$q$ region of the $\Delta$-HRD, our inferred mass ratios better follow the expected HRD offsets.

\begin{figure}
    \centering
    \includegraphics[width=1\linewidth]{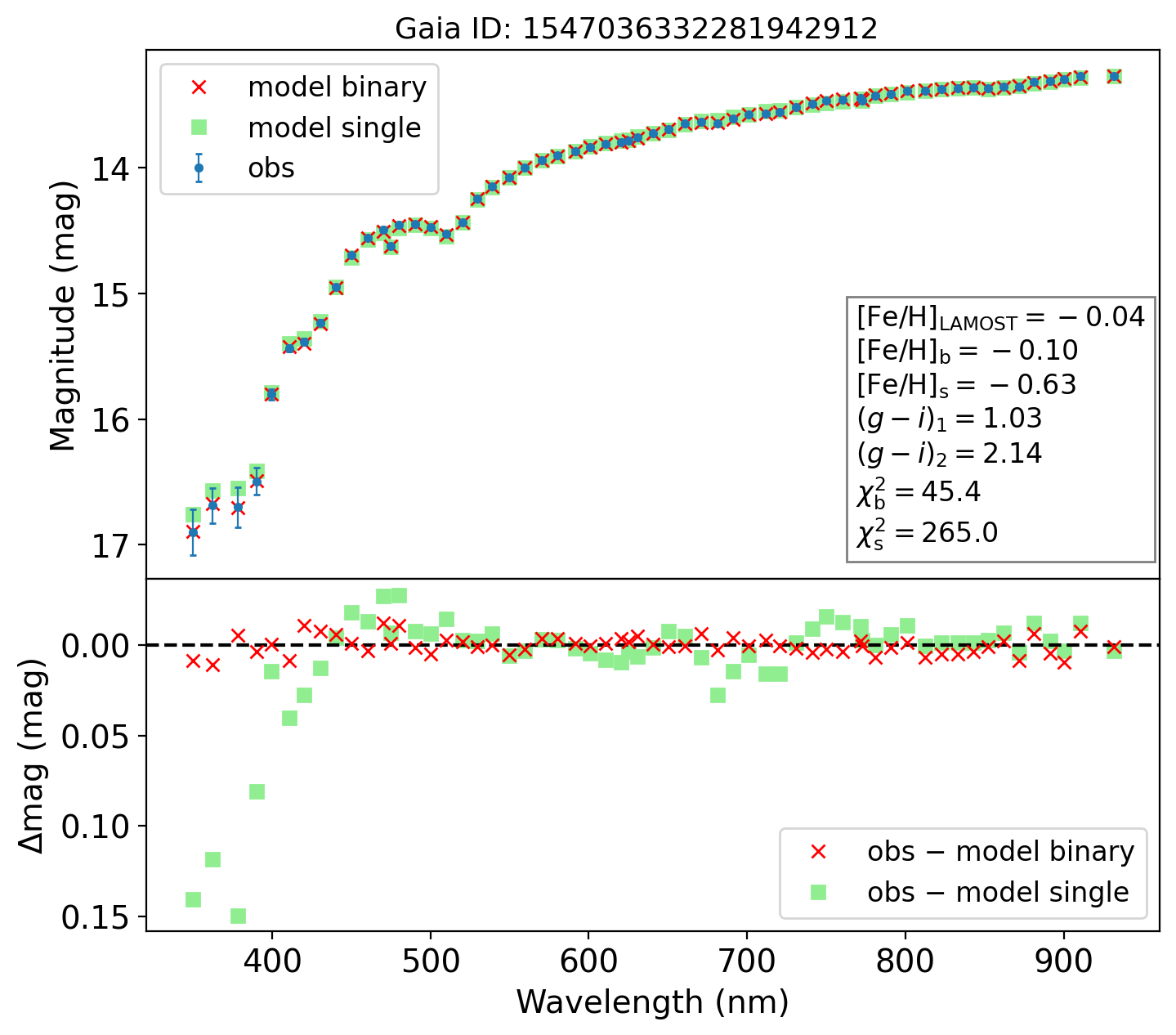}
    \caption{Observed SED and best-fit models. Upper panel: photometry (blue circles), binary-star model (red crosses), and single-star model (green squares). Lower panel: residuals $\Delta\mathrm{mag}\equiv m_{\mathrm{obs}}-m_{\mathrm{mod}}$ for both models. The inset reports the LAMOST and fitted metallicities, $\chi^2$, and the primary/secondary colors $(g-i)$.}
    \label{fig:lm_example}
\end{figure}

\begin{figure*}
    \centering
    \includegraphics[width=1\linewidth]{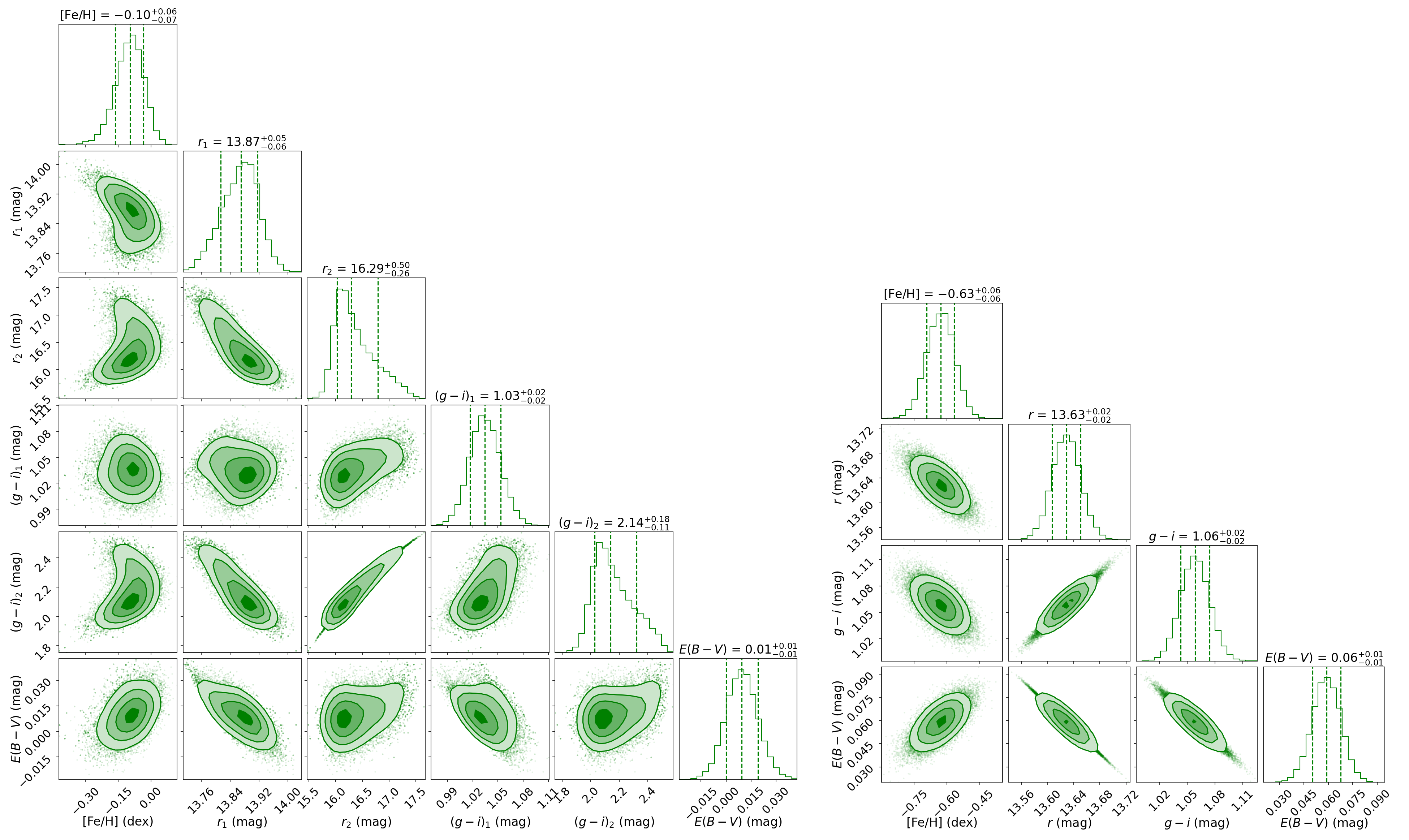}
    \caption{Posterior distributions for the illustrative LAMOST source shown in Figure~\ref{fig:lm_example}. Left: posterior distributions from the unresolved-binary fit, with the fitted parameters $\mathrm{[Fe/H]}$, $r_1$, $r_2$, $(g-i)_1$, $(g-i)_2$, and $E(B-V)$. Right: posterior distributions from the single-star fit, with the fitted parameters $\mathrm{[Fe/H]}$, $r$, $(g-i)$, and $E(B-V)$. The diagonal panels show the one-dimensional marginalized posteriors, while the off-diagonal panels show the corresponding two-dimensional marginalized posteriors. The vertical dashed lines indicate the 16th, 50th, and 84th percentiles of each posterior distribution.}
    \label{fig:corner_combined}
\end{figure*}

Figure~\ref{fig:lm_example} presents an illustrative case in which the binary-star model provides a visibly better match to the observed SED than the single-star model. The corresponding posterior distributions for the binary-star and single-star fits are shown in Figures~\ref{fig:corner_combined}, respectively. For this source, the binary-star model is strongly preferred, supporting an interpretation as an unresolved binary.

\subsection{Test with Cluster Data} \label{Test by Cluster}

We further assess the internal consistency of our parameter estimates using dwarf members of the open cluster M67 (NGC~2682), which are expected to share a common metallicity and reddening \citep{2021ApJ...923..129J}. This intrinsic uniformity provides a useful test bed for evaluating our method. We therefore restrict the sample to dwarfs that pass the corrected XP-spectra flag and have acceptable values of $phot\_bp\_rp\_excess\_factor$.

\begin{figure*}
    \centering
    \includegraphics[width=1\linewidth]{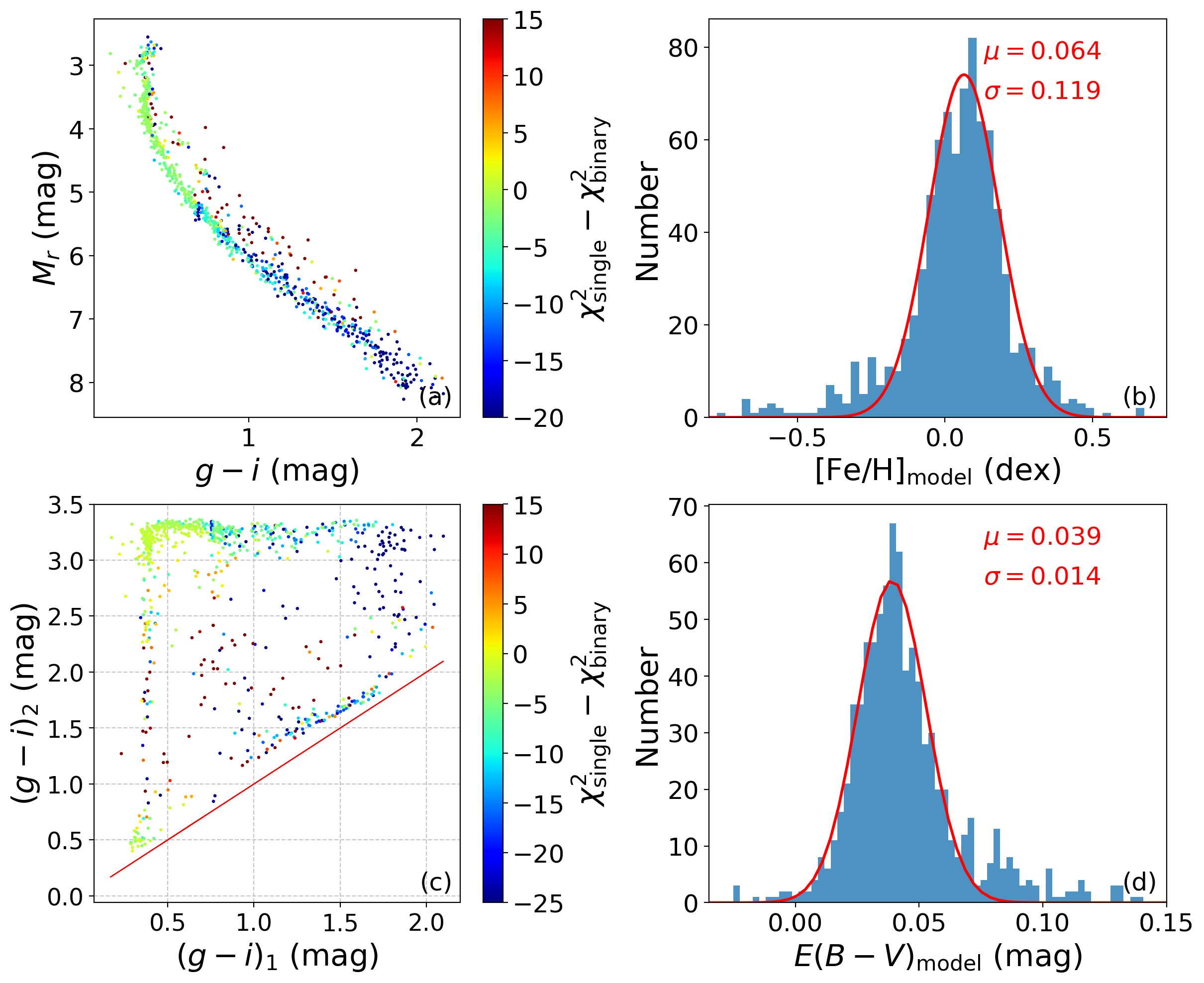}
    \caption{Upper left: M67 members on the HRD, colored by \(\Delta\chi^2\). Lower left: Primary and secondary \((g-i)\) colors from the binary-star model, colored by \(\Delta\chi^2\). Upper right: \([\mathrm{Fe/H}]\) from the best-fit model, with the Gaussian fit in red. Lower right: \(E(B-V)\) from the best-fit model, with the Gaussian fit in red.
}
    \label{fig:M67feh}
\end{figure*}

We fitted these sources with models that treat \(E(B-V)\) as a free parameter. Figure~\ref{fig:M67feh} shows that stars along the binary sequence in the HRD exhibit systematically larger $\Delta\chi^2$ than those on the single-star sequence, consistent with unresolved binarity. The resulting \([\mathrm{Fe/H}]\) and \(E(B-V)\) distributions are narrow: Gaussian fits give \(\mu=0.064\) and \(\sigma=0.119\)~dex for \([\mathrm{Fe/H}]\), and \(\mu=0.039\) and \(\sigma=0.014\)~mag for \(E(B-V)\). Overall, the small scatter among members supports the internal consistency of our inferred cluster parameters.

\subsection{Test with Gaia Non-single-star Solutions} \label{Test by Gaia}

Gaia DR3 provides a large sample of non-single-star systems via the Non-Single Star (NSS) catalog \citep{2023A&A...674A..34G}, which compiles binaries and multiples identified using several detection and modeling approaches.
For binaries with well-constrained orbits, dynamical masses for both components are available in the \texttt{binary\_masses} table in the Gaia Archive; our mass comparisons use this table.
We select sources with measured secondary masses, require the corrected XP-spectra flag and acceptable $phot\_bp\_rp\_excess\_factor$ values, and exclude special objects and multiple systems following \citet{2023A&A...674A..34G}, retaining only main-sequence--main-sequence binary solutions.

From this quality-selected \texttt{binary\_masses} sample, we randomly selected 1{,}200 sources with $(g-i)>0$ for the comparison analysis. Because many of these objects lie at low Galactic latitude, where the SFD extinction map is unreliable, we fit each source with both the single-star and binary-star models while treating $E(B-V)$ as a free parameter.

\begin{figure*}
    \centering
    \includegraphics[width=1\linewidth]{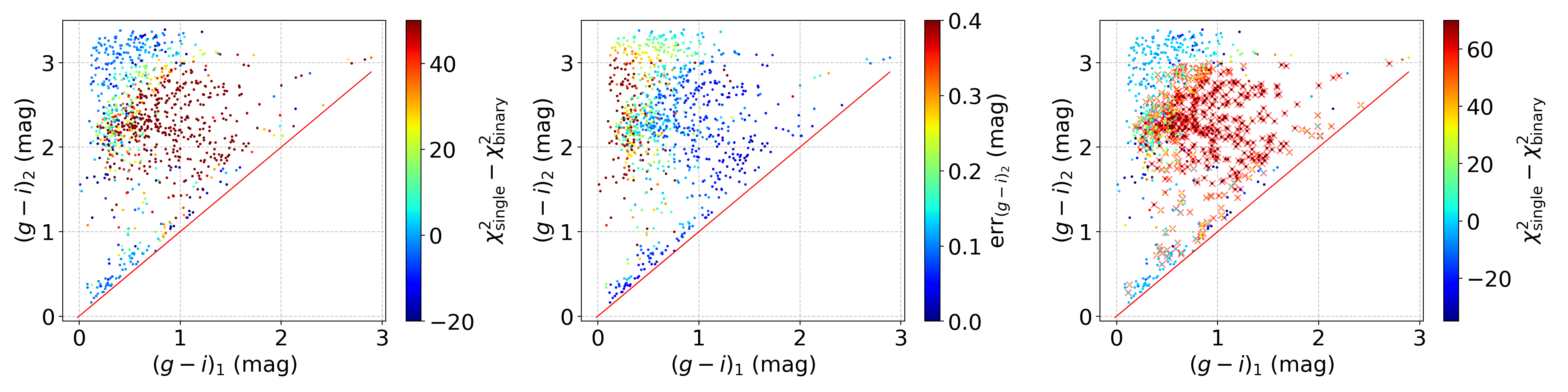}
    \caption{Primary versus secondary $(g-i)$ colors inferred from the binary-star model. Points are colored by $\Delta\chi^2$ (left) and by the secondary color uncertainty (middle). The right panel highlights the selected sources (red crosses).}
    \label{fig:gaia_ebv_ccd_p&s}
\end{figure*}

The resulting primary--secondary color distributions and the final selection are shown in Figure~\ref{fig:gaia_ebv_ccd_p&s}. To define a comparison sample that our models can reproduce, we exclude sources with $\chi^2>200$ in both the single-star and binary-star fits, and then require $\mathrm{err}_{(g-i)_2}<0.2$ and $\Delta\chi^2>0$ (i.e., systems better fit by the binary-star model). The final selection is shown as red crosses in the right panel of Figure~\ref{fig:gaia_ebv_ccd_p&s}.

\begin{figure*}
    \centering
    \includegraphics[width=1\linewidth]{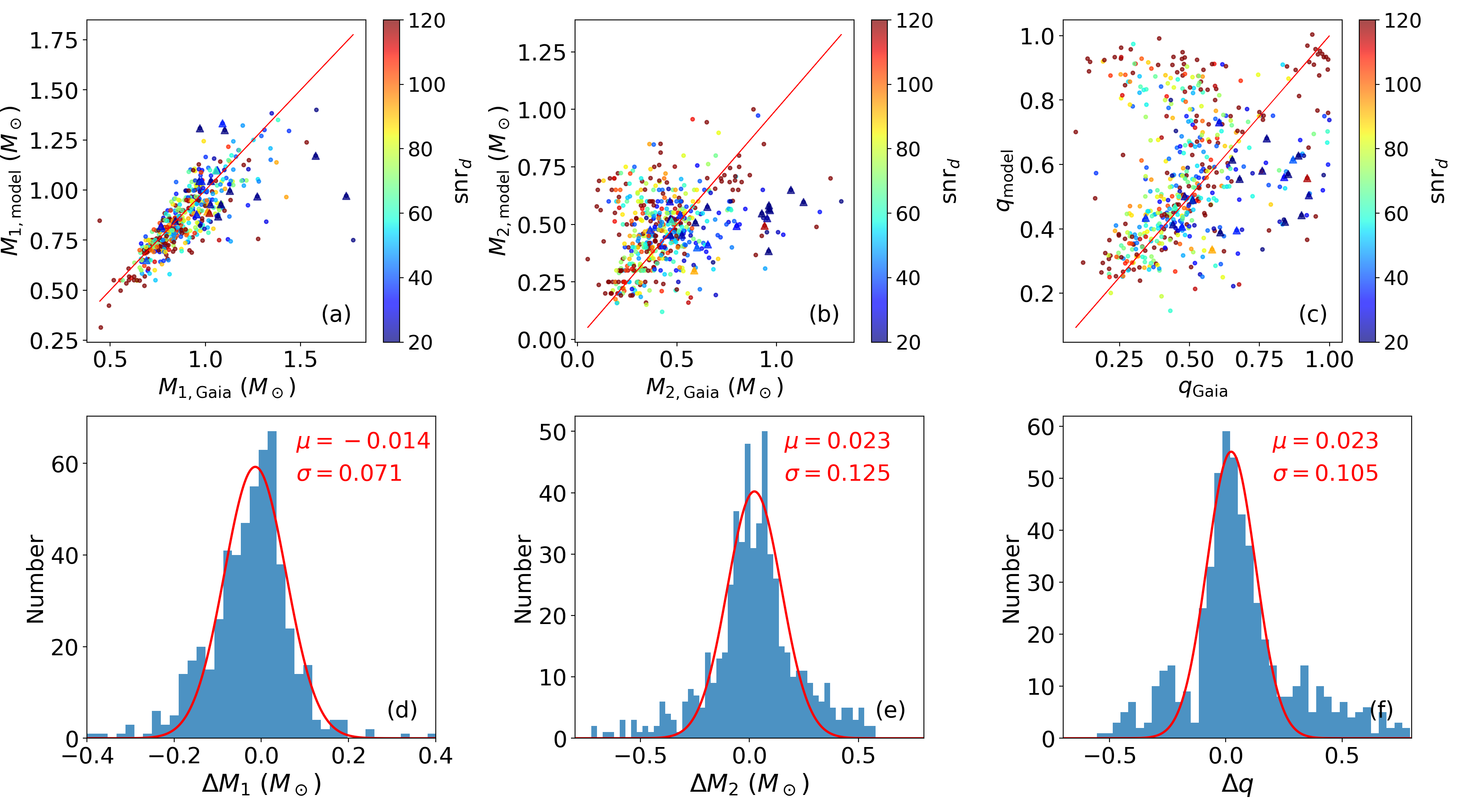}
    \caption{Comparison of primary mass, secondary mass, and mass ratio inferred by our models with the Gaia NSS values, for the selected Gaia test sources (red crosses in Figure~\ref{fig:gaia_ebv_ccd_p&s}). Triangles mark sources with $\Delta M_r<-1$\,mag, where $\Delta M_r\equiv M_r^{\mathrm{obs}}-M_r^{\mathrm{mod}}$ is the offset from the single-star color--absolute-magnitude relation; we interpret these as systems whose primaries lie near or beyond the MSTO. The red lines show the one-to-one relation. The bottom panels show the residual distributions (model minus Gaia) for each quantity, with Gaussian fits shown in red.}
    \label{fig:gaia_ebv_mass&q}
\end{figure*}

Figure~\ref{fig:gaia_ebv_mass&q} compares our inferred component masses with the Gaia NSS values; the color scale encodes the distance signal-to-noise ratio, $\mathrm{snr}_d\equiv d/\sigma_d$. Triangles denote sources with $\Delta M_r<-1$, which we interpret as systems whose primaries are near or beyond the MSTO. The primary-mass estimates agree well overall, whereas larger deviations appear for some secondary masses and mass ratios. Gaussian fits to the residuals yield mean offsets of $-0.014$, $0.023$, and $0.023$, with scatters of $0.071\,M_\odot$, $0.125\,M_\odot$, and $0.105$ for the primary mass, secondary mass, and mass ratio, respectively.

Apart from the upper-left group in Figure~\ref{fig:gaia_ebv_mass&q}(c), which we discuss separately below, these discrepancies occur mainly in systems with low $\mathrm{snr}_d$ and for sources near the MSTO. 
For low-$\mathrm{snr}_d$ sources, distance errors propagate directly into $M_r$ and therefore into the inferred luminosity excess assigned to the companion. Near the MSTO, our binary model already relaxes the main-sequence constraint for blue primaries with $(g-i)_1<0.75$, permitting primaries brighter than the main sequence (Section~\ref{subsec:Model Construction}). In this regime, an excess above the main sequence can be produced either by evolutionary brightening of the primary or by the companion’s flux, and these contributions are largely degenerate. Because our stellar loci depend only on color and $\mathrm{[Fe/H]}$, the SED offers limited leverage to break this degeneracy, degrading constraints on $M_2$ and $q$. Moreover, our color--absolute-magnitude relation is age-independent and therefore cannot capture the intrinsic HRD broadening near the MSTO, where stars of different ages and metallicities overlap. In future work, we will incorporate stellar age to better separate these effects and improve constraints for MSTO systems.

\begin{figure}
    \centering
    \includegraphics[width=1\linewidth]{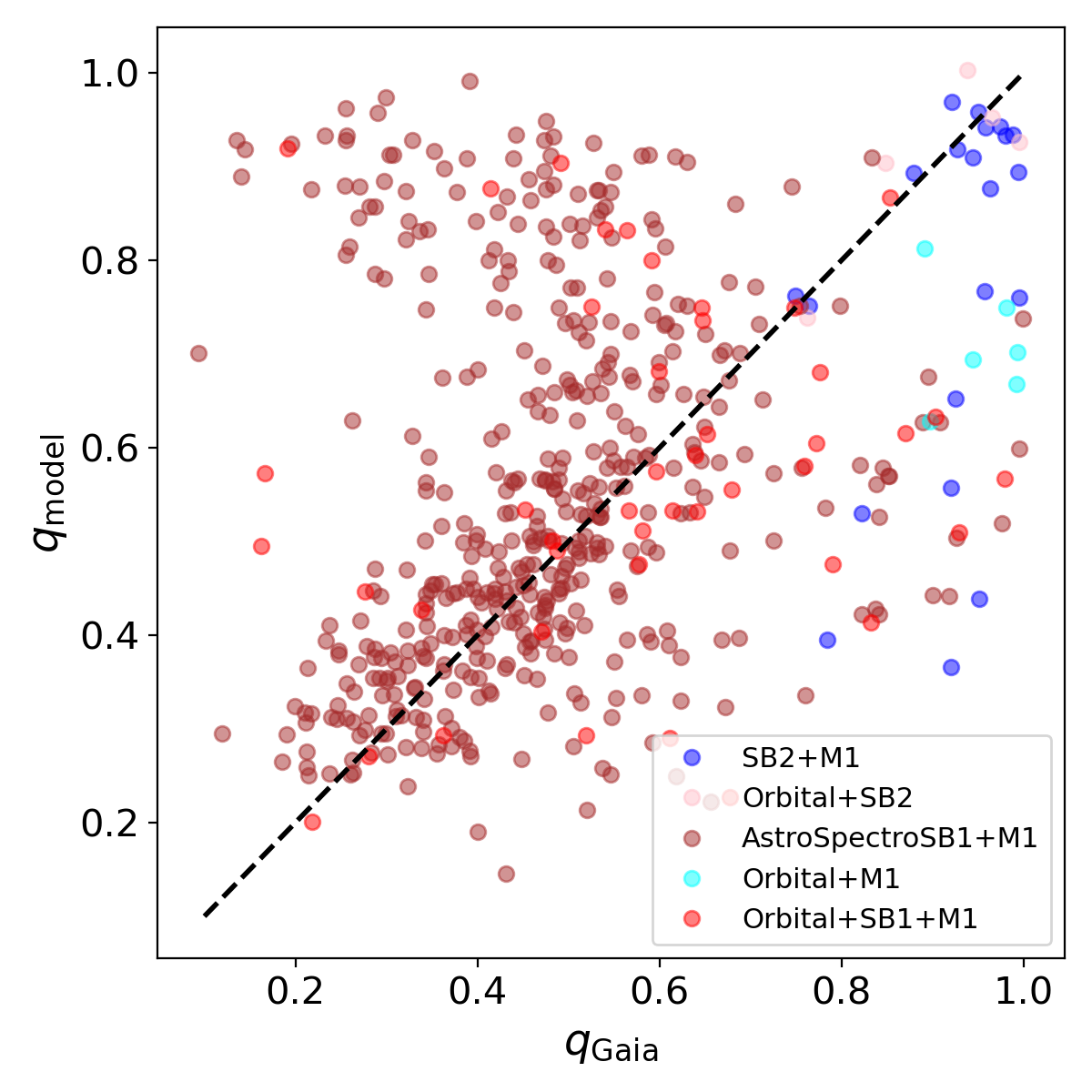}
    \caption{Mass-ratio comparison between our model and Gaia measurements. Points are colored by the Gaia solution type (see legend).}
    \label{fig:mass_ratio}
\end{figure}

Sources in the upper-left region of Figure~\ref{fig:gaia_ebv_mass&q}(c) show the largest disagreements with Gaia mass-ratio estimates. 
{All of these systems have Gaia solutions based on single-lined spectroscopic (SB1) orbits (\texttt{AstroSpectroSB1+M1} or \texttt{Orbital+SB1+M1}; Figure~\ref{fig:mass_ratio}); none of the SB2-based solutions lies in this region. In these solutions, \citet{2023A&A...674A..34G} derive $M_2$ from the spectroscopic mass function, $f(M)=M_2^3\sin^3 i/(M_1+M_2)^2\propto K_1^3$, with the inclination taken from the astrometric orbit. This approach assumes that the measured radial-velocity curve traces the primary alone, which holds only if the companion contributes negligible light in the RVS band. In near-equal-mass main-sequence pairs, however, the components have similar spectra, and for orbital periods of hundreds of days their velocity separation is comparable to, or smaller than, the Gaia RVS resolution element ($R\approx11\,500$; \citealt{2018A&A...616A...5C}). Their lines therefore remain blended over most of the orbit, so the system is not recognised as double-lined and is fitted as an SB1. The blended lines shift the measured velocity from that of the primary towards a weighted mean of the two components (the ``spectrocentre''), reducing the measured semi-amplitude $K_1$; for similar spectra, both $K_1$ and the inferred mass function can be severely underestimated \citep{2025A&A...693A.124G,ElBadry2024}. Because $f(M)\propto K_1^3$, even a modest reduction in $K_1$ can substantially bias $M_2$ (and thus $q$) low, with the effect strengthening as the companion approaches the primary in brightness. The photocentre orbit is reduced in the same way by the companion’s light, so the Gaia solution can remain internally consistent with a nearly dark companion and the bias is not flagged by the consistency check.}

\begin{figure}
    \centering
    \includegraphics[width=1\linewidth]{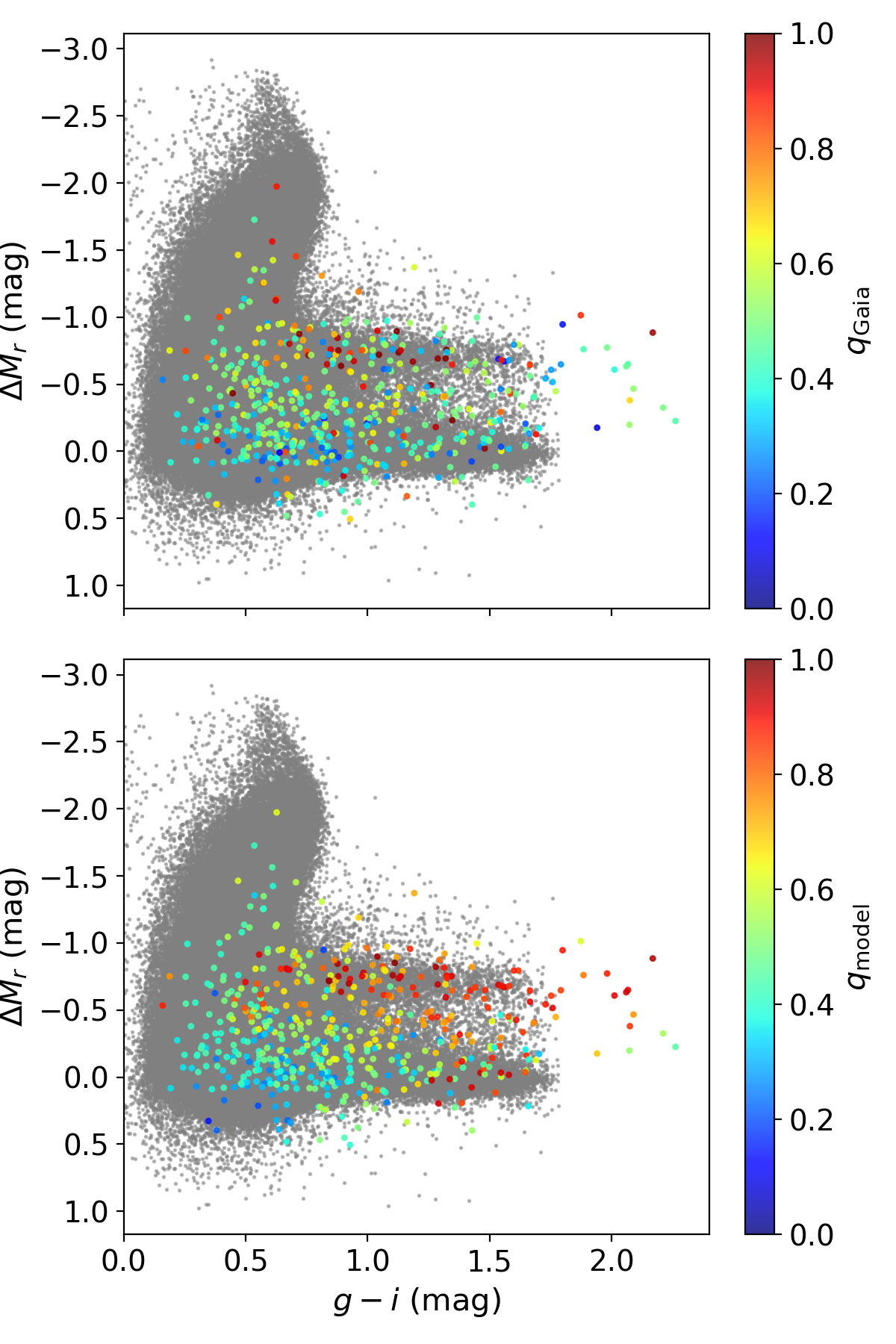}
    \caption{Distribution of the selected Gaia test sources on the $\Delta$-HRD. Colors encode the mass ratio: Gaia NSS \texttt{binary\_masses} values in the upper panel and our inferred values in the lower panel.}
    \label{fig:gaia_q_hrd}
\end{figure}

To further test the robustness of our measurements in this regime, we perform an additional consistency check based on each system’s $\Delta$-HRD position. Figure~\ref{fig:gaia_q_hrd} shows that the high mass-ratio main-sequence binaries selected by our method exhibit mass-ratio estimates that are broadly consistent with their offsets in the HR diagram. By contrast, the Gaia mass ratios display a weaker correspondence with $\Delta$-HRD. 
This behavior is expected if Gaia underestimates the mass ratios of these systems. They lie well above the single-star sequence in the $\Delta$-HRD, which requires a companion contributing a substantial fraction of the total light; by contrast, a main-sequence companion with $q\lesssim0.4$ would contribute only a few per cent and leave the system close to the single-star sequence. Because most of these sources have high $\mathrm{snr}_d$ (Figure~\ref{fig:gaia_ebv_mass&q}c), distance errors cannot explain the offset. We therefore attribute the discrepancy primarily to dilution in the single-lined radial velocities used by the Gaia solutions, rather than to a failure of our photometric decomposition.

\begin{figure}
    \centering
    \includegraphics[width=1\linewidth]{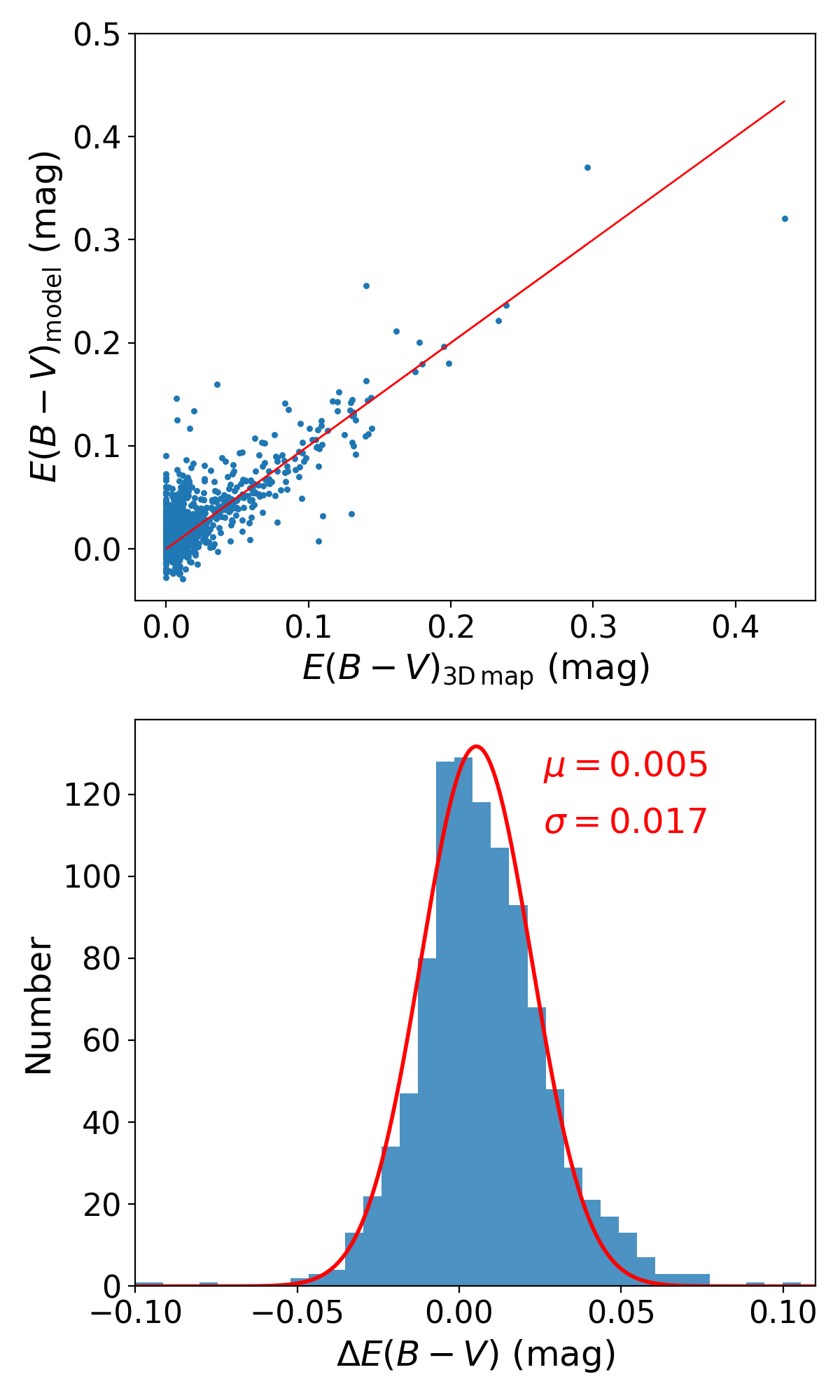}
    \caption{Comparison of $E(B-V)$ inferred by our models for the Gaia test sources with the values from the 3D extinction map of \citet{Wang2025}.}
    \label{fig:gaia_ebv_ebv}
\end{figure}

We also compare our extinction estimates to the independent three-dimensional dust map of \citet{Wang2025}. Figure~\ref{fig:gaia_ebv_ebv} shows excellent agreement between the two, with a scatter of 0.017\,mag, providing external validation for the robustness of our extinction determinations. 
Because many of these sources lie at low Galactic latitude---well outside the low-extinction regime of the training sample---this agreement suggests that the framework remains applicable, at least in part, to more heavily reddened stars. However, we adopt a single average Milky Way extinction curve (Appendix~\ref{app:extinction-coeffs}); spatial variations in the extinction law could introduce wavelength-dependent residuals that bias the inferred metallicities and companion parameters. In future work, we will incorporate spatially varying three-dimensional extinction laws to improve the extinction treatment and further refine the model.

\section{SUMMARY} \label{sec:SUMMARY}

We present a UV-to-IR SED-fitting framework to identify unresolved main-sequence binaries and infer component parameters from 65-band photometry. Our inputs combine synthetic miniJPAS magnitudes derived from corrected Gaia DR3 XP spectra with 2MASS and WISE measurements. We construct forward models for single stars and unresolved binaries using LAMOST-calibrated, metallicity-dependent stellar loci together with a dwarf-star color--absolute-magnitude relation. By fitting both model families with MCMC and using Gaia distances as HR-diagram constraints, we infer metallicity, reddening, and component photometric parameters; we then map these to component masses and mass ratios using PARSEC models. To avoid trapping the binary fits in local minima, we run several independent MCMC ensembles per source and adopt the one with the lowest $\chi^2$; the masses and mass ratios carry asymmetric uncertainties that reflect the shapes of the parameter posteriors.

Tests on simulated binaries demonstrate that the method reliably recovers binary parameters. Application to the LAMOST comparison sample indicates that the method is sensitive to binaries with $q\gtrsim0.4$. For this sample, the preferred model yields metallicities consistent with spectroscopy (mean offset $0.00$ dex; scatter 0.11 dex). Allowing extinction to vary freely, the inferred $E(B-V)$ values agree with the SFD map in the low-extinction regime with a scatter of 0.02 mag. Comparisons to an external Gaia XP--based binary catalog provide additional validation: classifications are broadly consistent for common sources, while our metallicities more closely follow the LAMOST spectroscopic scale and, at intermediate $q$, our mass-ratio estimates better match the expected $\Delta$-HRD offsets.

We further validate internal consistency using dwarf members of M67, which show narrow inferred metallicity and reddening distributions. Comparison with Gaia NSS solutions provides an external check on inferred masses: after quality cuts, our primary-mass estimates agree with Gaia NSS with a residual scatter of 0.071 $M_\odot$. For sources with discrepant solutions, the $\Delta$-HRD offsets indicate that our secondary-mass and mass-ratio estimates are often more consistent with the photometric constraints than the Gaia NSS values, with most discrepancies arising for low distance signal-to-noise sources or objects near the main-sequence turnoff.

The current framework has several limitations. Because our color--absolute-magnitude relation is age-independent, it is difficult to constrain primary and companion parameters near the main-sequence turnoff; moreover, we adopt a single average Milky Way extinction curve. In future work, we will incorporate stellar age and adopt spatially varying three-dimensional extinction laws.

Applied to large photometric data sets, this framework offers a scalable path to constructing binary-star catalogs with well-constrained component parameters, enabling population-level measurements of binary fractions and mass-ratio distributions and targeted studies of unusual systems.

\begin{acknowledgments}
This work is supported by the National Key Basic R\&D Program of China via 2024YFA1611601 and 2024YFA1611901, the National Natural Science Foundation of China through the projects NSFC 12222301, 12303039, 124B2055, and 12403024.  
This work is also supported by the China Manned Space Program with grant no. CMS-CSST-2025-A13.
This work has made use of data products from Gaia, LAMOST, 2MASS, and WISE. Gaia is a mission of the European Space Agency (ESA) (\url{https://www.cosmos.esa.int/gaia}). The Gaia data have been processed by the Gaia Data Processing and Analysis Consortium (DPAC, \url{https://www.cosmos.esa.int/web/gaia/dpac/consortium}), with funding for DPAC provided by national institutions, in particular those participating in the Gaia Multilateral Agreement. This work made use of the data from LAMOST (Large Sky Area Multi-Object Fiber Spectroscopic Telescope, also known as the Guoshoujing Telescope) (https://cstr.cn/31118.02.LAMOST ). LAMOST is a Chinese national mega-science facility, operated by National Astronomical Observatories, Chinese Academy of Sciences.

\end{acknowledgments}

\clearpage

\clearpage
\appendix
\renewcommand{\thefigure}{A\arabic{figure}}
\renewcommand{\theHfigure}{appendix.A.\arabic{figure}}
\setcounter{figure}{0}
\section{Stellar-locus fitting coefficients}
\label{app:stellar-locus-coeffs} 
\begin{figure*}
    \centering
    \includegraphics[width=1\linewidth]{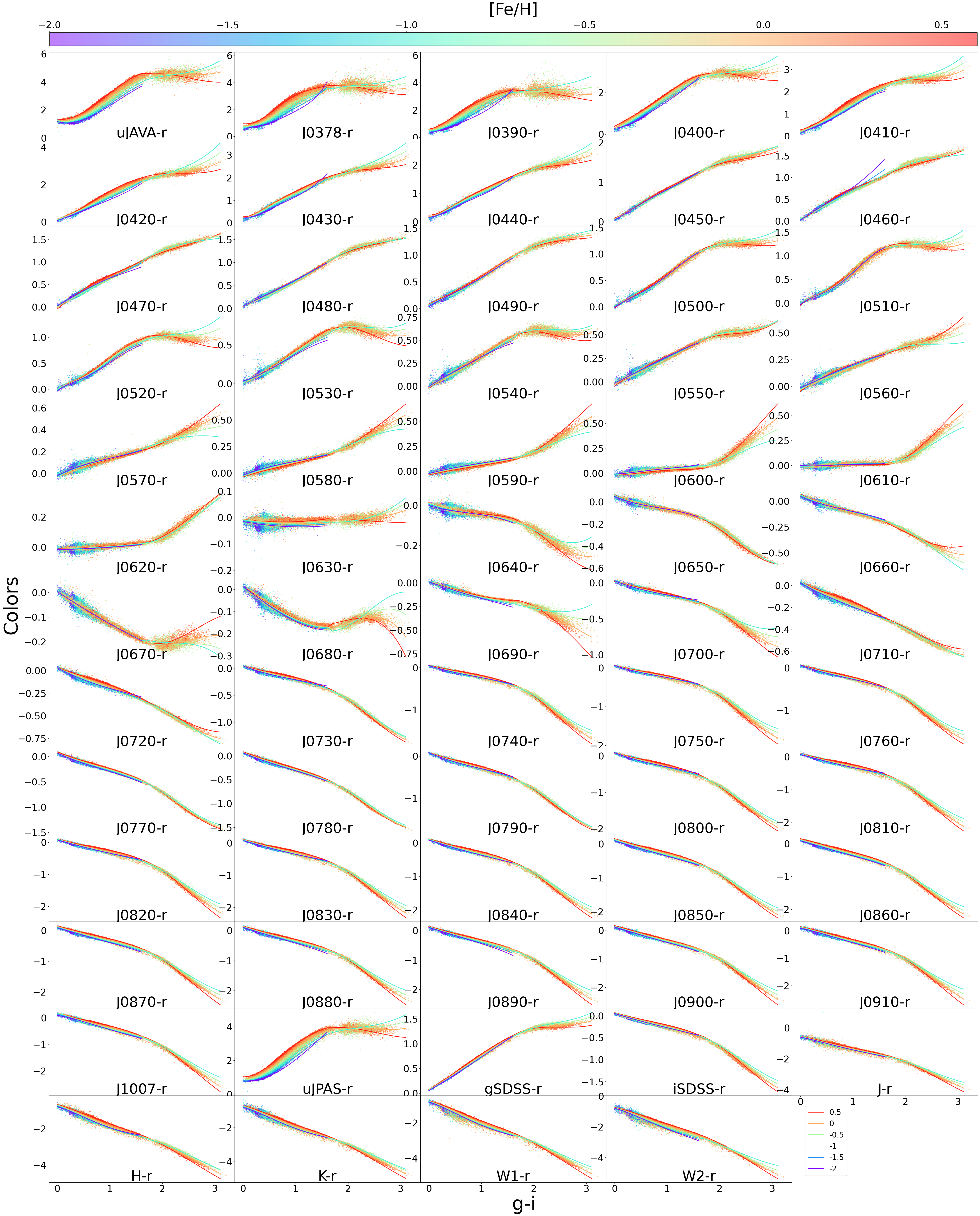}
    \caption{Metallicity-dependent stellar loci in multiple bandpasses. In each panel, dwarf stars are color-coded by $\mathrm{[Fe/H]}$, and the overplotted curves show the fitted loci at the labeled metallicities.}
    \label{fig:loci}
\end{figure*}

Figure~\ref{fig:loci} illustrates the fitted metallicity-dependent stellar loci in the individual bandpasses. The corresponding fitting coefficients are listed in the tables below: Stage~I coefficients appear in Tables~\ref{tab:table_2052_1}--\ref{tab:table_2063_all_new}, and Stage~II coefficients are given in Table~\ref{tab:table_righe}.

\section{Selection of single-star training samples}
\label{app:single-star-selection}

\begin{figure*}
    \centering
    \includegraphics[width=1\linewidth]{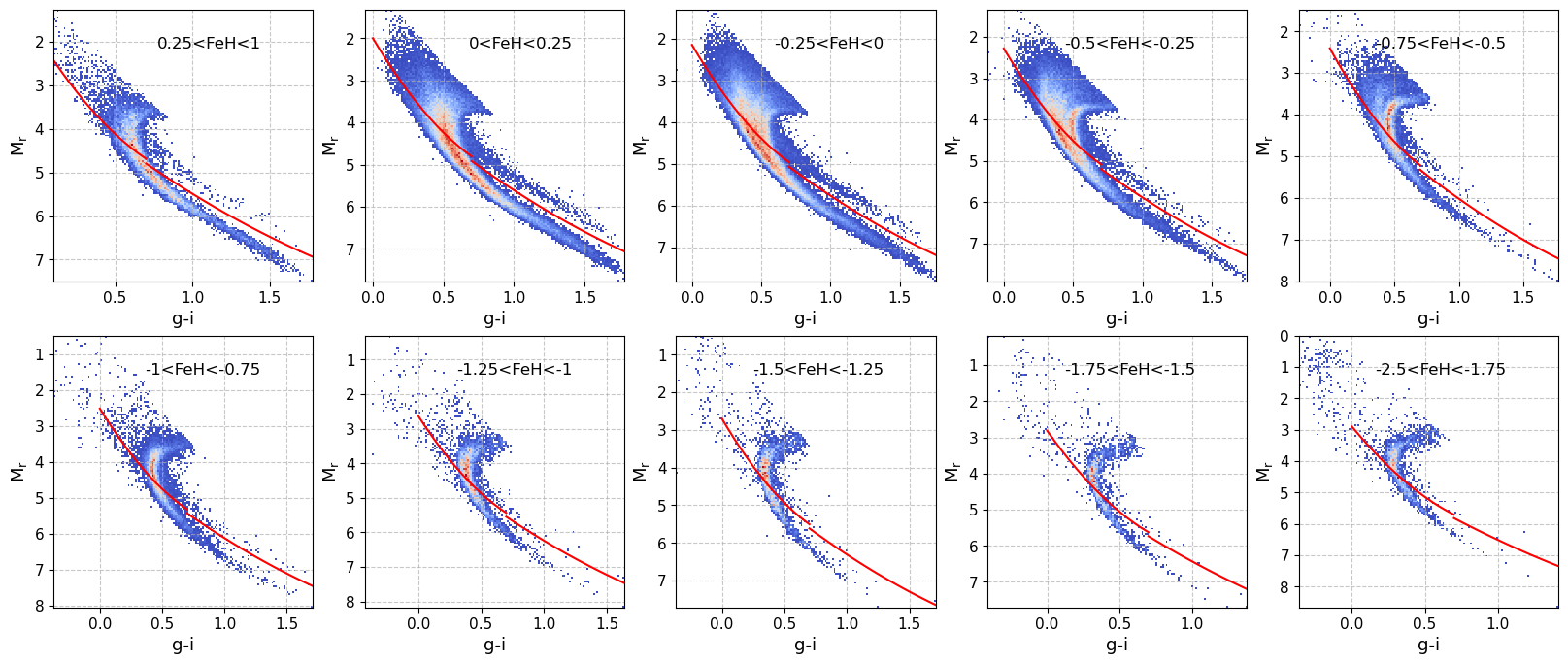}
    \caption{Selection of single-star AFGK dwarfs used to construct the color--absolute-magnitude relation. Each panel shows the density distribution in the HRD for the indicated metallicity interval. Sources above the red boundary, including candidate binaries and main-sequence turnoff stars, are excluded from the fit.}
    \label{fig:cutb_afgk}
\end{figure*}

\begin{figure}
    \centering
    \includegraphics[width=1\linewidth]{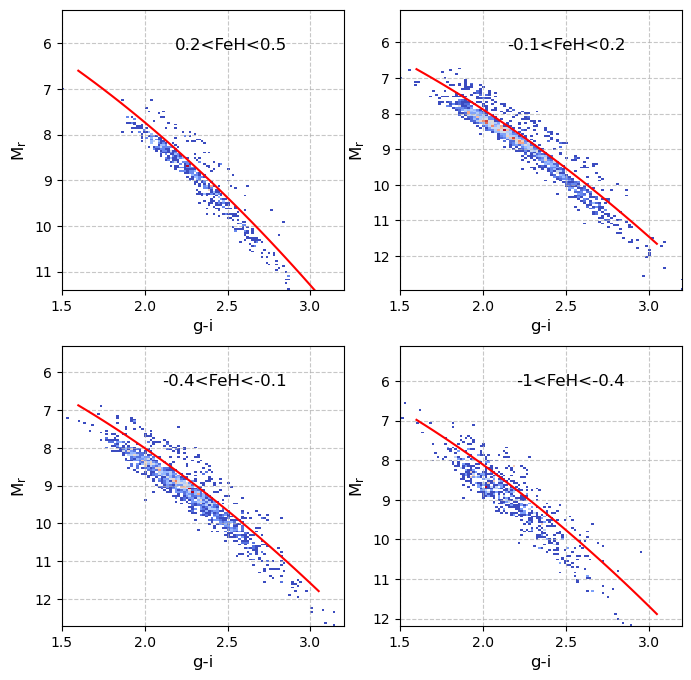}
    \caption{Selection of single-star M dwarfs used to construct the color--absolute-magnitude relation. Each panel shows the density distribution in the HRD for the indicated metallicity interval. Candidate binaries above the red boundary are excluded from the fit.}
    \label{fig:cutb_md}
\end{figure}

To construct the single-star color--absolute-magnitude relation, we removed unresolved-binary candidates and evolved stars, which would otherwise bias the sequence brightward and inflate its scatter. We performed the selection in $\mathrm{[Fe/H]}$ bins by excluding sources above an empirical boundary in the color--absolute-magnitude diagram. For AFGK stars, this boundary removes both the elevated binary sequence and the main-sequence turnoff (Figure~\ref{fig:cutb_afgk}); for M dwarfs, it primarily removes sources above the single-star sequence (Figure~\ref{fig:cutb_md}). We then combined the cleaned AFGK and M-dwarf samples to fit the metallicity-dependent relation described in Section~\ref{subsec:Metallicity-Dependent Color-Absolute Magnitude Relation}.


\section{Extinction (reddening) coefficients} 
\label{app:extinction-coeffs} 

The \citet{2024ApJ...971..127Z} toolkit\footnote{\url{https://github.com/vnohhf/XP_Extinction_Toolkit}} computes band-specific reddening coefficients from a representative Milky Way extinction curve. Given a filter transmission curve, stellar parameters ($T_{\mathrm{eff}},\log g$), and $E(B-V)$, it returns the corresponding coefficients. We use the response curves of miniJPAS, 2MASS, and WISE and, for our main-sequence-focused application, adopt coefficients evaluated at $T_{\mathrm{eff}}=5500\,\mathrm{K}$, $\log g=4.5$, and $E(B-V)=0.15\,\mathrm{mag}$ (Table~\ref{tab:kx_ebv015_teff5500_logg45}).

\section{Stellar-mass determination} 
\label{app:mass-determination} 
We infer stellar masses using PARSEC stellar-evolution models. We extract isochrones spanning $[\mathrm{Fe/H}]=-2.1$ to $+0.5$ in 0.1~dex steps and $\log(\mathrm{age/yr})=8.8$ to 10.1 in 0.1~dex steps, retaining only main-sequence points (PARSEC evolutionary stage \texttt{label}=1).
For each binary component, we compute the absolute $r$-band magnitude $M_r$ from the apparent magnitude and the distance, and form the observed vector $\{M_r,(g-i),[\mathrm{Fe/H}]\}$ 
from the posterior medians, with asymmetric credible intervals defined by the 16th and 84th percentiles of the corresponding posteriors. 
We then project this measurement onto the model HR diagram and assign the mass of the model point that minimizes the error-normalized distance.
The lower and upper limits of the mass are taken as the minimum and maximum masses of all model points that fall within the credible-interval volume. 
If no model points fall within this volume, we iteratively expand the bounds until at least one point is included. For redder stars, where the isochrone locus is largely insensitive to age, we instead determine the mass using only $\{(g-i),[\mathrm{Fe/H}]\}$ and their credible intervals.
We define the mass ratio as $q\equiv M_2/M_1$, computed from the component masses above; its bounds follow from the corresponding mass bounds. Because we do not assume symmetric credible intervals, the inferred uncertainties on the masses and mass ratios are generally asymmetric and reflect the skewness of the underlying posteriors.

\clearpage

\renewcommand{\thetable}{A\arabic{table}}
\setcounter{table}{0}

\begin{onecolumngrid}
\begin{center}
    \begin{deluxetable}{rrrrrrrr}
\tabletypesize{\scriptsize}
\tablecaption{Coefficients for stellar loci Stage\,I \label{tab:table_2052_1}}
\tablewidth{3pt}
\tablehead{
\colhead{Colors} & \colhead{\(a_{1}\)} & \colhead{\(a_{2}\)} & \colhead{\(a_{3}\)} & \colhead{\(a_{4}\)} & \colhead{\(a_{5}\)} & \colhead{\(a_{6}\)} & \colhead{\(a_{7}\)} 
}
\startdata
\(uJAVA - r\) & \(1.293e+00\) & \(1.062e-01\) & \(-8.515e-01\) & \(2.008e-02\) & \(3.218e-01\) & \(3.976e+00\) & \(9.639e-02\) \\
\(J0390 - r\) & \(5.779e-01\) & \(7.099e-02\) & \(-4.685e-02\) & \(-3.174e-02\) & \(8.676e-01\) & \(5.458e+00\) & \(6.329e-01\) \\
\(J0440 - r\) & \(1.800e-01\) & \(8.208e-02\) & \(1.935e-01\) & \(3.492e-02\) & \(-2.644e-01\) & \(2.304e+00\) & \(-1.372e-01\) \\
\(J0450 - r\) & \(7.767e-02\) & \(2.794e-02\) & \(5.655e-01\) & \(5.173e-03\) & \(-1.461e-01\) & \(7.613e-01\) & \(-2.519e-02\) \\
\(J0500 - r\) & \(-5.822e-03\) & \(-2.308e-02\) & \(6.950e-01\) & \(-1.292e-02\) & \(1.320e-01\) & \(-6.262e-01\) & \(7.214e-02\) \\
\(J0510 - r\) & \(-3.281e-02\) & \(-1.941e-02\) & \(9.202e-01\) & \(-1.009e-02\) & \(8.506e-02\) & \(-1.620e+00\) & \(5.217e-02\) \\
\(J0520 - r\) & \(-1.377e-02\) & \(-5.404e-03\) & \(7.267e-01\) & \(-9.169e-03\) & \(5.384e-02\) & \(-1.289e+00\) & \(6.354e-02\) \\
\(J0530 - r\) & \(2.212e-02\) & \(7.839e-03\) & \(2.191e-01\) & \(7.120e-03\) & \(-1.497e-02\) & \(2.708e-01\) & \(-1.175e-02\) \\
\(J0540 - r\) & \(-1.714e-02\) & \(-1.228e-02\) & \(3.965e-01\) & \(-2.927e-03\) & \(4.271e-02\) & \(-2.010e-01\) & \(2.214e-02\) \\
\(J0550 - r\) & \(-2.832e-02\) & \(-2.497e-02\) & \(3.306e-01\) & \(-8.630e-03\) & \(1.085e-01\) & \(-9.053e-02\) & \(4.072e-02\) \\
\(J0560 - r\) & \(-3.854e-02\) & \(-1.881e-02\) & \(3.283e-01\) & \(-2.868e-03\) & \(1.042e-01\) & \(-1.387e-01\) & \(2.165e-02\) \\
\(J0570 - r\) & \(-3.337e-02\) & \(-6.644e-04\) & \(2.559e-01\) & \(3.175e-03\) & \(-8.612e-03\) & \(-1.323e-01\) & \(-5.040e-03\) \\
\(J0580 - r\) & \(-3.331e-02\) & \(-7.217e-03\) & \(1.744e-01\) & \(-1.800e-04\) & \(-1.567e-02\) & \(-6.682e-02\) & \(-3.495e-03\) \\
\(J0590 - r\) & \(-3.263e-02\) & \(-9.107e-03\) & \(1.265e-01\) & \(-2.506e-03\) & \(-8.912e-03\) & \(-4.540e-02\) & \(-2.283e-04\) \\
\(J0600 - r\) & \(-1.499e-02\) & \(-2.636e-03\) & \(1.506e-02\) & \(3.613e-04\) & \(-1.886e-02\) & \(1.242e-01\) & \(-8.154e-03\) \\
\(J0610 - r\) & \(-7.422e-03\) & \(1.597e-03\) & \(-2.663e-02\) & \(2.911e-03\) & \(-8.847e-03\) & \(1.438e-01\) & \(-9.628e-03\) \\
\(J0620 - r\) & \(-1.300e-02\) & \(1.308e-03\) & \(1.017e-02\) & \(-3.974e-04\) & \(1.479e-03\) & \(8.592e-03\) & - \\
\(J0630 - r\) & \(-7.293e-03\) & \(1.058e-03\) & \(-2.805e-02\) & \(-1.428e-03\) & \(4.559e-03\) & \(1.691e-02\) & - \\
\(J0640 - r\) & \(1.365e-02\) & \(2.781e-03\) & \(-1.158e-01\) & \(-2.387e-03\) & \(-9.245e-03\) & \(8.979e-02\) & \(3.539e-03\) \\
\(J0650 - r\) & \(6.141e-02\) & \(5.019e-03\) & \(-2.251e-01\) & \(-9.714e-04\) & \(-1.009e-02\) & \(1.553e-01\) & \(8.241e-04\) \\
\(J0660 - r\) & \(6.474e-02\) & \(6.050e-03\) & \(-2.573e-01\) & \(-2.447e-03\) & \(-1.113e-02\) & \(1.306e-01\) & \(3.929e-03\) \\
\(J0670 - r\) & \(4.515e-03\) & \(3.731e-04\) & \(-1.416e-01\) & \(-4.937e-05\) & \(3.106e-04\) & \(1.123e-02\) & - \\
\(J0680 - r\) & \(1.070e-02\) & \(2.726e-03\) & \(-1.609e-01\) & \(1.909e-03\) & \(3.197e-03\) & \(-4.176e-03\) & \(-3.441e-03\) \\
\(J0690 - r\) & \(1.040e-02\) & \(-3.747e-03\) & \(-1.533e-01\) & \(-2.498e-03\) & \(3.691e-02\) & \(-7.882e-02\) & \(1.317e-02\) \\
\(J0700 - r\) & \(2.342e-02\) & \(-4.380e-03\) & \(-2.345e-01\) & \(-1.269e-03\) & \(5.085e-02\) & \(8.508e-02\) & \(1.078e-02\) \\
\(J0710 - r\) & \(3.864e-02\) & \(7.597e-04\) & \(-3.176e-01\) & \(-3.866e-03\) & \(3.301e-02\) & \(2.331e-01\) & \(2.484e-02\) \\
\(J0720 - r\) & \(3.269e-02\) & \(4.637e-04\) & \(-2.445e-01\) & \(2.139e-04\) & \(6.499e-02\) & \(6.522e-02\) & \(1.225e-02\) \\
\(J0730 - r\) & \(4.056e-02\) & \(-7.423e-04\) & \(-2.920e-01\) & \(4.285e-04\) & \(7.792e-02\) & \(1.137e-01\) & \(1.464e-02\) \\
\(J0740 - r\) & \(5.193e-02\) & \(4.304e-03\) & \(-3.300e-01\) & \(2.379e-03\) & \(6.490e-02\) & \(1.381e-01\) & \(9.741e-03\) \\
\(J0750 - r\) & \(6.324e-02\) & \(1.382e-03\) & \(-3.852e-01\) & \(7.150e-04\) & \(6.440e-02\) & \(1.858e-01\) & \(1.096e-02\) \\
\(J0760 - r\) & \(6.635e-02\) & \(6.907e-03\) & \(-3.990e-01\) & \(3.914e-03\) & \(5.109e-02\) & \(1.811e-01\) & \(3.206e-03\) \\
\(J0770 - r\) & \(8.101e-02\) & \(1.964e-02\) & \(-4.385e-01\) & \(7.578e-03\) & \(1.588e-02\) & \(1.848e-01\) & \(-7.436e-03\) \\
\(J0780 - r\) & \(7.965e-02\) & \(1.701e-02\) & \(-4.350e-01\) & \(6.029e-03\) & \(2.102e-02\) & \(1.399e-01\) & \(-2.670e-03\) \\
\(J0790 - r\) & \(8.316e-02\) & \(-2.067e-03\) & \(-4.678e-01\) & \(-3.418e-04\) & \(8.436e-02\) & \(2.014e-01\) & \(1.685e-02\) \\
\(J0800 - r\) & \(7.940e-02\) & \(-8.959e-03\) & \(-4.476e-01\) & \(-2.785e-03\) & \(1.150e-01\) & \(1.678e-01\) & \(2.589e-02\) \\
\(J0810 - r\) & \(9.240e-02\) & \(6.464e-03\) & \(-4.693e-01\) & \(-4.148e-03\) & \(3.592e-02\) & \(1.410e-01\) & \(3.315e-02\) \\
\(J0820 - r\) & \(1.070e-01\) & \(6.172e-03\) & \(-5.173e-01\) & \(-5.020e-03\) & \(5.058e-02\) & \(2.121e-01\) & \(4.036e-02\) \\
\(J0830 - r\) & \(1.113e-01\) & \(1.255e-02\) & \(-5.026e-01\) & \(-3.177e-03\) & \(3.197e-02\) & \(1.353e-01\) & \(3.210e-02\) \\
\(J0840 - r\) & \(1.228e-01\) & \(1.287e-02\) & \(-5.339e-01\) & \(-2.601e-03\) & \(4.421e-02\) & \(1.541e-01\) & \(3.538e-02\) \\
\(J0850 - r\) & \(1.432e-01\) & \(2.177e-02\) & \(-5.376e-01\) & \(-3.808e-03\) & \(4.764e-02\) & \(1.436e-01\) & \(4.495e-02\) \\
\(J0860 - r\) & \(1.347e-01\) & \(3.395e-02\) & \(-4.285e-01\) & \(4.910e-03\) & \(4.419e-02\) & \(-7.837e-02\) & \(2.080e-02\) \\
\(J0870 - r\) & \(1.422e-01\) & \(3.063e-02\) & \(-4.977e-01\) & \(2.469e-03\) & \(4.325e-02\) & \(2.294e-02\) & \(2.859e-02\) \\
\(J0880 - r\) & \(1.425e-01\) & \(2.054e-02\) & \(-5.628e-01\) & \(2.921e-04\) & \(4.815e-02\) & \(1.070e-01\) & \(3.357e-02\) \\
\(J0890 - r\) & \(1.417e-01\) & \(1.370e-02\) & \(-5.747e-01\) & \(-1.291e-03\) & \(7.614e-02\) & \(8.878e-02\) & \(4.181e-02\) \\
\(J0910 - r\) & \(1.363e-01\) & \(1.785e-02\) & \(-4.304e-01\) & \(-3.809e-03\) & \(5.877e-02\) & \(-2.563e-01\) & \(4.845e-02\) \\
\(J1007 - r\) & \(1.966e-01\) & \(1.467e-04\) & \(-6.195e-01\) & \(-2.469e-02\) & \(7.457e-02\) & \(1.233e-02\) & \(1.188e-01\) \\
\(gSDSS - r\) & \(6.918e-02\) & \(7.311e-03\) & \(6.157e-01\) & \(-1.109e-03\) & \(3.610e-02\) & \(1.420e-01\) & \(2.113e-02\) \\
\(iSDSS - r\) & \(6.918e-02\) & \(7.311e-03\) & \(-3.843e-01\) & \(-1.109e-03\) & \(3.610e-02\) & \(1.420e-01\) & \(2.113e-02\) \\
\(J - r\) & \(-5.587e-01\) & \(7.266e-03\) & \(-6.757e-01\) & \(-8.820e-03\) & \(1.521e-01\) & \(-4.298e-01\) & \(8.582e-02\) \\
\(H - r\) & \(-6.840e-01\) & \(3.618e-02\) & \(-9.011e-01\) & \(-8.589e-03\) & \(1.897e-01\) & \(-7.610e-01\) & \(1.140e-01\) \\
\(K_s - r\) & \(-7.226e-01\) & \(2.836e-02\) & \(-8.690e-01\) & \(-9.850e-03\) & \(1.914e-01\) & \(-9.621e-01\) & \(1.125e-01\) \\
\(W1 - r\) & \(-7.679e-01\) & \(1.904e-02\) & \(-8.524e-01\) & \(-5.581e-03\) & \(2.332e-01\) & \(-1.038e+00\) & \(1.073e-01\) \\
\(W2 - r\) & \(-7.615e-01\) & \(2.201e-02\) & \(-7.737e-01\) & \(1.824e-03\) & \(2.731e-01\) & \(-1.056e+00\) & \(1.013e-01\) \\
\enddata

\tablenotetext{}{%
\footnotesize
Fitting coefficients for Stage\,I of the stellar loci. \(f(x, y) = a_{1} \cdot x^5 + a_{2} \cdot x^4 \cdot y + a_{3} \cdot x^3 \cdot y^2 + a_{4} \cdot x^4 + a_{5} \cdot x^3 \cdot y + a_{6} \cdot x^2 \cdot y^2 + a_{7} \cdot x^3 + a_{8} \cdot x^2 \cdot y + a_{9} \cdot x \cdot y^2 + a_{10} \cdot x^2 + a_{11} \cdot x \cdot y + a_{12} \cdot y^2 + a_{13} \cdot x + a_{14} \cdot y + a_{15}\), where \(x \equiv g - i\) and \(y \equiv [\text{Fe/H}]\).
}
\end{deluxetable}

    \begin{deluxetable}{rrrrrrrrr}
\tabletypesize{\scriptsize}
\tablecaption{Coefficients for stellar loci Stage\,I \label{tab:table_2052_2}}
\tablewidth{3pt}
\tablehead{
\colhead{Colors} & \colhead{\(a_{8}\)} & \colhead{\(a_{9}\)} & \colhead{\(a_{10}\)} & \colhead{\(a_{11}\)} & \colhead{\(a_{12}\)} & \colhead{\(a_{13}\)} & \colhead{\(a_{14}\)} & \colhead{\(a_{15}\)} 
}
\startdata
\(uJAVA - r\) & \(2.054e-01\) & \(-1.575e+00\) & \(-4.970e-02\) & \(-1.744e-01\) & \(1.063e-01\) & - & - & - \\
\(J0390 - r\) & \(-5.435e-01\) & \(-4.990e+00\) & \(-1.241e+00\) & \(-3.691e-01\) & \(1.889e+00\) & \(5.710e-01\) & \(2.510e-01\) & \(-2.935e-01\) \\
\(J0440 - r\) & \(3.616e-01\) & \(-2.743e+00\) & \(1.155e-01\) & \(-1.028e-01\) & \(1.405e+00\) & \(-1.747e-02\) & \(-1.475e-02\) & \(-2.656e-01\) \\
\(J0450 - r\) & \(3.056e-01\) & \(-1.018e+00\) & \(6.526e-02\) & \(-1.748e-01\) & \(5.380e-01\) & \(-2.890e-02\) & \(2.384e-02\) & \(-9.957e-02\) \\
\(J0500 - r\) & \(-8.421e-02\) & \(8.725e-01\) & \(-6.330e-02\) & \(-1.643e-02\) & \(-3.618e-01\) & \(1.263e-02\) & \(1.301e-02\) & \(3.153e-02\) \\
\(J0510 - r\) & \(1.300e-02\) & \(2.525e+00\) & \(-5.096e-02\) & \(-8.762e-02\) & \(-1.353e+00\) & \(1.242e-02\) & \(2.828e-02\) & \(2.269e-01\) \\
\(J0520 - r\) & \(-8.250e-03\) & \(2.319e+00\) & \(-1.002e-01\) & \(-3.239e-02\) & \(-1.453e+00\) & \(3.813e-02\) & \(1.575e-02\) & \(2.970e-01\) \\
\(J0530 - r\) & \(2.012e-02\) & \(-9.542e-02\) & - & - & - & - & - & - \\
\(J0540 - r\) & \(-3.544e-02\) & \(2.117e-01\) & \(-1.901e-02\) & \(9.262e-03\) & \(-6.609e-02\) & - & - & - \\
\(J0550 - r\) & \(-9.570e-02\) & \(5.528e-02\) & \(-2.285e-02\) & \(2.399e-02\) & \(-1.092e-02\) & - & - & - \\
\(J0560 - r\) & \(-1.190e-01\) & \(3.085e-02\) & \(-1.146e-02\) & \(4.039e-02\) & \(7.283e-03\) & - & - & - \\
\(J0570 - r\) & \(3.141e-03\) & \(4.405e-02\) & - & - & - & - & - & - \\
\(J0580 - r\) & \(9.614e-03\) & \(2.465e-02\) & - & - & - & - & - & - \\
\(J0590 - r\) & \(4.660e-03\) & \(1.638e-02\) & - & - & - & - & - & - \\
\(J0600 - r\) & \(1.515e-02\) & \(-1.244e-01\) & \(4.277e-03\) & \(-6.233e-03\) & \(3.633e-02\) & - & - & - \\
\(J0610 - r\) & \(1.173e-03\) & \(-1.306e-01\) & \(1.845e-03\) & \(-1.304e-03\) & \(3.619e-02\) & - & - & - \\
\(J0620 - r\) & - & - & - & - & - & - & - & - \\
\(J0630 - r\) & - & - & - & - & - & - & - & - \\
\(J0640 - r\) & \(1.012e-02\) & \(-3.213e-02\) & - & - & - & - & - & - \\
\(J0650 - r\) & \(4.975e-03\) & \(-5.733e-02\) & - & - & - & - & - & - \\
\(J0660 - r\) & \(4.604e-03\) & \(-4.432e-02\) & - & - & - & - & - & - \\
\(J0670 - r\) & - & - & - & - & - & - & - & - \\
\(J0680 - r\) & \(-3.074e-03\) & \(2.139e-02\) & - & - & - & - & - & - \\
\(J0690 - r\) & \(-3.573e-02\) & \(1.010e-01\) & \(-1.044e-02\) & \(1.243e-02\) & \(-2.657e-02\) & - & - & - \\
\(J0700 - r\) & \(-2.480e-02\) & \(-3.495e-02\) & - & - & - & - & - & - \\
\(J0710 - r\) & \(2.463e-02\) & \(-1.566e-01\) & \(-1.151e-02\) & \(-2.681e-02\) & \(3.216e-02\) & - & - & - \\
\(J0720 - r\) & \(-3.173e-02\) & \(-3.206e-02\) & - & - & - & - & - & - \\
\(J0730 - r\) & \(-3.939e-02\) & \(-5.954e-02\) & - & - & - & - & - & - \\
\(J0740 - r\) & \(-3.180e-02\) & \(-7.061e-02\) & - & - & - & - & - & - \\
\(J0750 - r\) & \(-3.130e-02\) & \(-8.836e-02\) & - & - & - & - & - & - \\
\(J0760 - r\) & \(-2.806e-02\) & \(-8.653e-02\) & - & - & - & - & - & - \\
\(J0770 - r\) & \(-1.245e-02\) & \(-7.913e-02\) & - & - & - & - & - & - \\
\(J0780 - r\) & \(-1.293e-02\) & \(-6.016e-02\) & - & - & - & - & - & - \\
\(J0790 - r\) & \(-3.802e-02\) & \(-9.095e-02\) & - & - & - & - & - & - \\
\(J0800 - r\) & \(-4.981e-02\) & \(-8.261e-02\) & - & - & - & - & - & - \\
\(J0810 - r\) & \(8.192e-02\) & \(-2.910e-02\) & \(-5.816e-03\) & \(-5.685e-02\) & \(-2.067e-02\) & - & - & - \\
\(J0820 - r\) & \(6.409e-02\) & \(-8.855e-02\) & \(-1.168e-02\) & \(-5.063e-02\) & \(-3.396e-03\) & - & - & - \\
\(J0830 - r\) & \(7.606e-02\) & \(-1.358e-02\) & \(-8.411e-03\) & \(-5.367e-02\) & \(-2.658e-02\) & - & - & - \\
\(J0840 - r\) & \(5.453e-02\) & \(-1.369e-02\) & \(-1.648e-02\) & \(-4.728e-02\) & \(-2.908e-02\) & - & - & - \\
\(J0850 - r\) & \(3.621e-02\) & \(-7.322e-03\) & \(-2.781e-02\) & \(-4.093e-02\) & \(-3.094e-02\) & - & - & - \\
\(J0860 - r\) & \(-5.987e-03\) & \(1.399e-01\) & \(-1.464e-02\) & \(-1.130e-02\) & \(-6.235e-02\) & - & - & - \\
\(J0870 - r\) & \(1.145e-02\) & \(8.096e-02\) & \(-2.220e-02\) & \(-2.351e-02\) & \(-5.158e-02\) & - & - & - \\
\(J0880 - r\) & \(1.167e-02\) & \(3.013e-02\) & \(-3.035e-02\) & \(-2.608e-02\) & \(-4.106e-02\) & - & - & - \\
\(J0890 - r\) & \(-3.465e-02\) & \(4.963e-02\) & \(-4.174e-02\) & \(-7.279e-03\) & \(-4.566e-02\) & - & - & - \\
\(J0910 - r\) & \(1.725e-02\) & \(2.937e-01\) & \(-3.363e-02\) & \(-3.066e-02\) & \(-1.027e-01\) & - & - & - \\
\(J1007 - r\) & \(1.174e-01\) & \(1.315e-01\) & \(-6.900e-02\) & \(-9.927e-02\) & \(-7.352e-02\) & - & - & - \\
\(gSDSS - r\) & \(2.626e-02\) & \(-5.522e-02\) & \(-8.737e-03\) & \(-2.781e-02\) & \(-5.935e-03\) & - & - & - \\
\(iSDSS - r\) & \(2.626e-02\) & \(-5.522e-02\) & \(-8.737e-03\) & \(-2.781e-02\) & \(-5.935e-03\) & - & - & - \\
\(J - r\) & \(-4.313e-02\) & \(4.438e-01\) & \(-4.790e-02\) & \(-2.110e-02\) & \(-1.319e-01\) & - & - & - \\
\(H - r\) & \(2.567e-02\) & \(6.868e-01\) & \(-4.198e-02\) & \(-7.373e-02\) & \(-1.755e-01\) & - & - & - \\
\(K_s - r\) & \(4.960e-03\) & \(8.454e-01\) & \(-5.236e-02\) & \(-6.231e-02\) & \(-2.180e-01\) & - & - & - \\
\(W1 - r\) & \(-3.946e-02\) & \(9.078e-01\) & \(-5.881e-02\) & \(-4.810e-02\) & \(-2.365e-01\) & - & - & - \\
\(W2 - r\) & \(-8.224e-02\) & \(9.212e-01\) & \(-6.643e-02\) & \(-2.141e-02\) & \(-2.498e-01\) & - & - & - \\
\enddata

\tablenotetext{}{%
\footnotesize
Continuation of Table~\ref{tab:table_2052_1}.
}

\end{deluxetable}

\end{center}

\begin{center}
    \begin{splitdeluxetable*}{lcccccccBccccccccBccccccc}
\tabletypesize{\scriptsize}
\tablecaption{Coefficients for stellar loci Stage\,I\label{tab:table_2063_all_new}}
\tablewidth{3pt}
\tablehead{
\colhead{Colors} &
\colhead{$a_{1}$} & \colhead{$a_{2}$} & \colhead{$a_{3}$} & \colhead{$a_{4}$} &
\colhead{$a_{5}$} & \colhead{$a_{6}$} & \colhead{$a_{7}$} & \colhead{$a_{8}$} &
\colhead{$a_{9}$} & \colhead{$a_{10}$} & \colhead{$a_{11}$} & \colhead{$a_{12}$} &
\colhead{$a_{13}$} & \colhead{$a_{14}$} & \colhead{$a_{15}$} & \colhead{$a_{16}$} &
\colhead{$a_{17}$} & \colhead{$a_{18}$} & \colhead{$a_{19}$} & \colhead{$a_{20}$} &
\colhead{$a_{21}$} & \colhead{$a_{22}$}\\
}
\startdata 
$J0378 - r$ & $7.816e-01$ & $2.601e-01$ & $3.942e-01$ & $9.220e-02$ & $-1.202e+00$ & $-7.659e-01$ & $2.232e-02$ & $-1.478e-01$ & $5.916e+00$ & $1.122e+01$ & $-4.216e-02$ & $1.022e+00$ & $-7.711e+00$ & $-1.525e+01$ & $1.363e-01$ & $-1.475e+00$ & $3.925e+00$ & $7.903e+00$ & $-1.053e-01$ & $5.392e-01$ & $-7.043e-01$ & $-1.470e+00$\\
$J0400 - r$ & $3.012e-01$ & $1.449e-01$ & $1.298e+00$ & $3.929e-02$ & $2.240e-01$ & $2.765e-01$ & $-1.467e-02$ & $-2.026e-02$ & $-8.499e-01$ & $1.197e-01$ & $6.377e-02$ & $-8.203e-02$ & $8.995e-01$ & $-7.546e-02$ & $-5.430e-02$ & $3.831e-02$ & $-2.873e-01$ & $-3.116e-02$ & - & - & - & -\\
$J0410 - r$ & $2.494e-01$ & $9.734e-02$ & $7.424e-01$ & $-3.553e-02$ & $-3.926e-01$ & $5.960e-01$ & $-2.876e-02$ & $1.125e-01$ & $1.365e+00$ & $4.132e-01$ & $1.033e-01$ & $2.059e-01$ & $-1.058e+00$ & $-6.453e-01$ & $-4.477e-02$ & $-1.534e-01$ & $2.293e-01$ & $1.759e-01$ & - & - & - & -\\
$J0420 - r$ & $1.644e-01$ & $1.953e-02$ & $6.605e-01$ & $-5.825e-02$ & $-1.590e-02$ & $1.441e+00$ & $-2.316e-02$ & $1.842e-01$ & $4.570e-01$ & $-9.499e-01$ & $5.980e-02$ & $-2.133e-02$ & $-2.454e-01$ & $1.833e-01$ & - & - & - & - & - & - & - & - \\
$J0430 - r$ & $2.337e-01$ & $8.294e-02$ & $8.955e-02$ & $-6.108e-03$ & $-1.843e-01$ & $3.515e+00$ & $-2.757e-02$ & $-1.090e-01$ & $8.869e-02$ & $-4.440e+00$ & $6.224e-02$ & $1.068e-01$ & $2.325e-01$ & $2.464e+00$ & $-4.783e-02$ & $-2.316e-02$ & $-1.332e-01$ & $-5.111e-01$ & - & - & - & -\\
$J0460 - r$ & $6.439e-02$ & $-1.107e-02$ & $3.574e-01$ & $-5.099e-02$ & $-6.278e-02$ & $1.250e+00$ & $-1.743e-02$ & $2.481e-01$ & $5.087e-01$ & $-1.723e+00$ & $8.618e-02$ & $-1.643e-01$ & $-5.240e-01$ & $9.236e-01$ & $-6.871e-02$ & $1.519e-02$ & $1.478e-01$ & $-1.710e-01$ & - & - & - & -\\
$J0470 - r$ & $-1.135e-02$ & $-5.642e-02$ & $7.688e-01$ & $-1.949e-02$ & $3.018e-01$ & $1.662e-02$ & $-1.537e-05$ & $1.685e-01$ & $-2.277e-01$ & $-2.223e-01$ & $2.152e-02$ & $-8.000e-02$ & $4.070e-02$ & $9.927e-02$ & - & - & - & - & - & - & - & -\\
$J0480 - r$ & $4.470e-02$ & $-1.041e-02$ & $5.925e-01$ & $8.401e-03$ & $8.998e-02$ & $-1.945e-01$ & $6.379e-03$ & $3.519e-02$ & $-8.847e-02$ & $2.017e-01$ & $-4.768e-03$ & $-2.617e-02$ & $2.507e-02$ & $-4.783e-02$ & - & - & - & - & - & - & - & -\\
$J0490 - r$ & $7.395e-02$ & $1.303e-02$ & $4.708e-01$ & $1.027e-02$ & $-9.147e-04$ & $-1.086e-01$ & $6.731e-03$ & $1.372e-02$ & $3.331e-02$ & $2.359e-01$ & $-9.406e-03$ & $-2.031e-02$ & $-2.256e-02$ & $-8.393e-02$ & - & - & - & - & - & - & - & -\\
$J0900 - r$ & $1.416e-01$ & $2.633e-02$ & $-5.063e-01$ & $2.088e-02$ & $4.337e-02$ & $-1.076e-01$ & $8.682e-03$ & $-6.716e-03$ & $-8.385e-03$ & $2.029e-01$ & $-1.402e-02$ & $-1.911e-02$ & $-1.237e-02$ & $-8.443e-02$ & - & - & - & - & - & - & - & -\\
$uJPAS - r$ & $9.741e-01$ & $1.321e-01$ & $-4.828e-01$ & $-5.181e-03$ & $2.942e-01$ & $4.253e+00$ & $-1.183e-02$ & $4.597e-01$ & $5.203e-01$ & $-2.319e+00$ & $1.040e-01$ & $-6.584e-01$ & $-8.295e-01$ & $3.178e-01$ & $-8.828e-02$ & $2.248e-01$ & $2.663e-01$ & $1.214e-02$ & - & - & - & - \\
\enddata
\tablenotetext{}{%
\footnotesize
Fitting coefficients for Stage\,I of the stellar loci. \(
f(x, y) = a_{1} \cdot x^6 + a_{2} \cdot x^5 \cdot y + a_{3} \cdot x^4 \cdot y^2 + a_{4} \cdot x^3 \cdot y^3 + a_{5} \cdot x^5 + a_{6} \cdot x^4 \cdot y + a_{7} \cdot x^3 \cdot y^2 + a_{8} \cdot x^2 \cdot y^3 + a_{9} \cdot x^4 + a_{10} \cdot x^3 \cdot y + a_{11} \cdot x^2 \cdot y^2 + a_{12} \cdot x \cdot y^3 + a_{13} \cdot x^3 + a_{14} \cdot x^2 \cdot y + a_{15} \cdot x \cdot y^2 + a_{16} \cdot y^3 + a_{17} \cdot x^2 + a_{18} \cdot x \cdot y + a_{19} \cdot y^2 + a_{20} \cdot x + a_{21} \cdot y + a_{22}
\), where \(x \equiv g - i\) and \(y \equiv [\text{Fe/H}]\).
}
\end{splitdeluxetable*}
\end{center}

\end{onecolumngrid} 

\begin{onecolumngrid}

\centerwidetable
\begin{deluxetable}{rrrrrrrrrrr}
\tabletypesize{\scriptsize}
\tablecaption{Coefficients for stellar loci Stage\,II \label{tab:table_righe}}
\setlength{\tabcolsep}{3pt} 
\tablehead{
\colhead{Colors} & \colhead{\(a_{1}\)} & \colhead{\(a_{2}\)} & \colhead{\(a_{3}\)} & \colhead{\(a_{4}\)} & \colhead{\(a_{5}\)} & \colhead{\(a_{6}\)} & \colhead{\(a_{7}\)} & \colhead{\(a_{8}\)} & \colhead{\(a_{9}\)} & \colhead{\(a_{10}\)} 
}
\startdata
\(uJAVA - r\) & \(-1.268e-01\) & \(-2.754e-01\) & \(-3.886e-01\) & \(6.902e-01\) & \(3.866e-01\) & \(7.896e-01\) & \(-4.940e+00\) & \(1.270e-01\) & \(1.162e+01\) & \(-4.427e+00\) \\
\(J0378 - r\) & \(-2.347e-01\) & \(-8.659e-03\) & \(-2.155e-01\) & \(5.800e-01\) & \(-1.360e-01\) & \(2.743e-01\) & \(-4.174e+00\) & \(4.184e-01\) & \(9.625e+00\) & \(-3.435e+00\) \\
\(J0390 - r\) & \(-3.086e-01\) & \(1.951e-01\) & \(-4.078e-01\) & \(3.436e-01\) & \(-4.741e-01\) & \(1.161e+00\) & \(-2.542e+00\) & \(-5.747e-01\) & \(5.937e+00\) & \(-1.073e+00\) \\
\(J0400 - r\) & \(-1.016e-01\) & \(-2.577e-02\) & \(-3.360e-01\) & \(5.487e-01\) & \(-3.795e-02\) & \(9.734e-01\) & \(-3.942e+00\) & \(-5.446e-01\) & \(9.288e+00\) & \(-4.263e+00\) \\
\(J0410 - r\) & \(-1.262e-02\) & \(-3.786e-02\) & \(-2.210e-01\) & \(5.858e-01\) & \(8.962e-02\) & \(5.098e-01\) & \(-4.065e+00\) & \(-1.051e-01\) & \(9.527e+00\) & \(-4.952e+00\) \\
\(J0420 - r\) & \(6.342e-02\) & \(3.692e-02\) & \(-3.837e-01\) & \(5.489e-01\) & \(2.684e-02\) & \(1.084e+00\) & \(-3.716e+00\) & \(-5.855e-01\) & \(8.721e+00\) & \(-4.451e+00\) \\
\(J0430 - r\) & \(3.928e-02\) & \(7.337e-02\) & \(-2.090e-01\) & \(2.915e-01\) & \(-1.003e-01\) & \(5.215e-01\) & \(-2.060e+00\) & \(-2.071e-01\) & \(5.273e+00\) & \(-2.383e+00\) \\
\(J0440 - r\) & \(5.020e-02\) & \(4.270e-03\) & \(-1.291e-01\) & \(2.866e-01\) & \(2.590e-02\) & \(2.906e-01\) & \(-2.067e+00\) & \(-7.804e-02\) & \(5.325e+00\) & \(-2.816e+00\) \\
\(J0450 - r\) & \(1.964e-02\) & \(-5.942e-02\) & \(3.269e-02\) & \(2.175e-01\) & \(8.958e-02\) & \(-2.567e-01\) & \(-1.633e+00\) & \(3.323e-01\) & \(4.363e+00\) & \(-2.438e+00\) \\
\(J0460 - r\) & \(3.492e-02\) & \(-6.190e-02\) & \(1.151e-01\) & \(1.447e-01\) & \(9.850e-02\) & \(-5.394e-01\) & \(-1.112e+00\) & \(5.546e-01\) & \(3.147e+00\) & \(-1.735e+00\) \\
\(J0470 - r\) & \(1.506e-02\) & \(-2.084e-02\) & \(1.202e-01\) & \(1.276e-01\) & \(3.686e-02\) & \(-5.342e-01\) & \(-9.959e-01\) & \(5.435e-01\) & \(2.898e+00\) & \(-1.601e+00\) \\
\(J0480 - r\) & \(2.214e-03\) & \(-1.092e-02\) & \(4.414e-02\) & \(1.322e-01\) & \(-6.284e-04\) & \(-2.087e-01\) & \(-1.089e+00\) & \(2.194e-01\) & \(3.211e+00\) & \(-1.878e+00\) \\
\(J0490 - r\) & \(-1.672e-02\) & \(2.633e-03\) & \(-1.826e-02\) & \(9.998e-02\) & \(-2.032e-02\) & \(1.642e-02\) & \(-8.635e-01\) & \(3.722e-02\) & \(2.613e+00\) & \(-1.410e+00\) \\
\(J0500 - r\) & \(5.772e-02\) & \(1.330e-02\) & \(-8.622e-02\) & \(2.567e-01\) & \(1.574e-02\) & \(2.688e-01\) & \(-1.909e+00\) & \(-2.076e-01\) & \(4.717e+00\) & \(-2.674e+00\) \\
\(J0510 - r\) & \(1.132e-01\) & \(-1.711e-03\) & \(-1.715e-01\) & \(3.254e-01\) & \(8.714e-02\) & \(6.014e-01\) & \(-2.314e+00\) & \(-5.344e-01\) & \(5.372e+00\) & \(-2.864e+00\) \\
\(J0520 - r\) & \(-5.483e-04\) & \(2.354e-02\) & \(-1.935e-01\) & \(2.914e-01\) & \(-3.641e-02\) & \(6.324e-01\) & \(-2.109e+00\) & \(-4.778e-01\) & \(4.937e+00\) & \(-2.742e+00\) \\
\(J0530 - r\) & \(-3.499e-02\) & \(-2.664e-02\) & \(-7.135e-02\) & \(2.433e-01\) & \(-3.907e-05\) & \(1.771e-01\) & \(-1.814e+00\) & \(-7.195e-02\) & \(4.316e+00\) & \(-2.581e+00\) \\
\(J0540 - r\) & \(-2.493e-02\) & \(-2.832e-02\) & \(-2.650e-02\) & \(1.965e-01\) & \(2.378e-02\) & \(3.823e-02\) & \(-1.446e+00\) & \(1.593e-02\) & \(3.457e+00\) & \(-2.093e+00\) \\
\(J0550 - r\) & \(4.356e-03\) & \(2.985e-03\) & \(3.074e-02\) & \(1.607e-01\) & \(-4.178e-03\) & \(-1.480e-01\) & \(-1.111e+00\) & \(1.580e-01\) & \(2.612e+00\) & \(-1.566e+00\) \\
\(J0560 - r\) & \(-2.705e-03\) & \(-4.274e-03\) & \(1.082e-01\) & \(8.756e-02\) & \(-8.922e-03\) & \(-1.369e-01\) & \(-5.713e-01\) & \(3.623e-01\) & \(1.367e+00\) & \(-7.751e-01\) \\
\(J0570 - r\) & \(-8.988e-03\) & \(6.149e-03\) & \(1.177e-01\) & \(-1.712e-02\) & \(-2.213e-02\) & \(-4.071e-01\) & \(1.504e-01\) & \(3.391e-01\) & \(-2.012e-01\) & \(2.230e-01\) \\
\(J0580 - r\) & \(-8.952e-04\) & \(2.748e-02\) & \(8.746e-02\) & \(-5.157e-02\) & \(-3.768e-02\) & \(-2.836e-01\) & \(3.995e-01\) & \(2.213e-01\) & \(-7.433e-01\) & \(5.509e-01\) \\
\(J0590 - r\) & \(1.289e-02\) & \(4.073e-02\) & \(9.692e-02\) & \(-6.714e-02\) & \(-5.615e-02\) & \(-3.052e-01\) & \(5.509e-01\) & \(2.266e-01\) & \(-1.152e+00\) & \(8.266e-01\) \\
\(J0600 - r\) & \(1.297e-02\) & \(2.527e-02\) & \(1.111e-01\) & \(-1.208e-01\) & \(-3.028e-02\) & \(-3.375e-01\) & \(9.747e-01\) & \(2.335e-01\) & \(-2.177e+00\) & \(1.533e+00\) \\
\(J0610 - r\) & \(-6.656e-03\) & \(1.346e-02\) & \(5.531e-02\) & \(-1.319e-01\) & \(-1.883e-02\) & \(-1.369e-01\) & \(1.066e+00\) & \(6.559e-02\) & \(-2.409e+00\) & \(1.681e+00\) \\
\(J0620 - r\) & \(-1.655e-02\) & \(-6.678e-03\) & \(-9.558e-03\) & \(-6.557e-02\) & \(8.152e-03\) & \(5.504e-02\) & \(5.390e-01\) & \(-5.509e-02\) & \(-1.193e+00\) & \(8.208e-01\) \\
\(J0630 - r\) & \(-1.686e-02\) & \(-2.124e-02\) & \(-3.814e-02\) & \(1.398e-02\) & \(2.838e-02\) & \(1.226e-01\) & \(-7.969e-02\) & \(-8.169e-02\) & \(1.613e-01\) & \(-1.230e-01\) \\
\(J0640 - r\) & \(-7.034e-03\) & \(-1.339e-02\) & \(-5.352e-02\) & \(8.543e-02\) & \(1.534e-02\) & \(1.665e-01\) & \(-5.751e-01\) & \(-1.193e-01\) & \(1.117e+00\) & \(-7.367e-01\) \\
\(J0650 - r\) & \(-1.440e-03\) & \(4.671e-04\) & \(2.920e-02\) & \(1.422e-01\) & \(-1.719e-02\) & \(-1.374e-01\) & \(-9.762e-01\) & \(1.389e-01\) & \(1.871e+00\) & \(-1.210e+00\) \\
\(J0660 - r\) & \(4.673e-04\) & \(1.621e-02\) & \(1.210e-01\) & \(9.209e-02\) & \(-3.824e-02\) & \(-4.536e-01\) & \(-5.976e-01\) & \(4.032e-01\) & \(1.017e+00\) & \(-6.690e-01\) \\
\(J0670 - r\) & \(2.256e-03\) & \(2.585e-02\) & \(6.391e-02\) & \(-6.739e-02\) & \(-3.215e-02\) & \(-2.113e-01\) & \(5.070e-01\) & \(1.728e-01\) & \(-1.215e+00\) & \(7.274e-01\) \\
\(J0680 - r\) & \(1.497e-02\) & \(1.009e-02\) & \(-1.259e-01\) & \(-1.235e-01\) & \(-5.095e-03\) & \(4.543e-01\) & \(7.561e-01\) & \(-3.915e-01\) & \(-1.470e+00\) & \(7.527e-01\) \\
\(J0690 - r\) & \(1.696e-02\) & \(-3.081e-02\) & \(-2.123e-01\) & \(7.857e-02\) & \(5.573e-02\) & \(7.169e-01\) & \(-6.310e-01\) & \(-5.839e-01\) & \(1.366e+00\) & \(-1.087e+00\) \\
\(J0700 - r\) & \(1.558e-02\) & \(-1.430e-02\) & \(-1.360e-01\) & \(1.711e-01\) & \(3.128e-02\) & \(4.406e-01\) & \(-1.222e+00\) & \(-3.447e-01\) & \(2.403e+00\) & \(-1.689e+00\) \\
\(J0710 - r\) & \(1.244e-02\) & \(3.060e-02\) & \(4.235e-02\) & \(5.515e-02\) & \(-3.083e-02\) & \(-1.630e-01\) & \(-3.482e-01\) & \(1.656e-01\) & \(4.709e-01\) & \(-3.893e-01\) \\
\(J0720 - r\) & \(8.075e-03\) & \(3.061e-02\) & \(7.402e-02\) & \(7.332e-02\) & \(-3.435e-02\) & \(-2.840e-01\) & \(-4.751e-01\) & \(2.803e-01\) & \(6.904e-01\) & \(-5.094e-01\) \\
\(J0730 - r\) & \(2.009e-02\) & \(-1.168e-02\) & \(4.442e-03\) & \(2.074e-01\) & \(2.929e-02\) & \(-8.540e-02\) & \(-1.495e+00\) & \(1.371e-01\) & \(2.837e+00\) & \(-1.935e+00\) \\
\(J0740 - r\) & \(2.782e-02\) & \(-1.648e-02\) & \(-5.757e-02\) & \(2.947e-01\) & \(2.926e-02\) & \(9.991e-02\) & \(-2.159e+00\) & \(-9.831e-04\) & \(4.249e+00\) & \(-2.883e+00\) \\
\(J0750 - r\) & \(2.230e-02\) & \(-1.821e-02\) & \(-7.771e-02\) & \(3.216e-01\) & \(2.574e-02\) & \(1.618e-01\) & \(-2.360e+00\) & \(-5.161e-02\) & \(4.667e+00\) & \(-3.173e+00\) \\
\(J0760 - r\) & \(2.416e-02\) & \(1.732e-05\) & \(-5.481e-02\) & \(2.871e-01\) & \(-2.199e-03\) & \(1.051e-01\) & \(-2.065e+00\) & \(-2.087e-02\) & \(3.970e+00\) & \(-2.696e+00\) \\
\(J0770 - r\) & \(9.722e-03\) & \(1.045e-02\) & \(7.156e-03\) & \(2.232e-01\) & \(-1.529e-02\) & \(-7.294e-02\) & \(-1.523e+00\) & \(1.063e-01\) & \(2.710e+00\) & \(-1.822e+00\) \\
\(J0780 - r\) & \(-3.789e-03\) & \(1.938e-02\) & \(3.477e-02\) & \(2.249e-01\) & \(-4.059e-02\) & \(-1.829e-01\) & \(-1.558e+00\) & \(2.157e-01\) & \(2.809e+00\) & \(-1.928e+00\) \\
\(J0790 - r\) & \(9.495e-03\) & \(-5.952e-03\) & \(4.041e-02\) & \(2.557e-01\) & \(5.122e-03\) & \(-2.409e-01\) & \(-1.822e+00\) & \(2.992e-01\) & \(3.385e+00\) & \(-2.314e+00\) \\
\(J0800 - r\) & \(1.152e-02\) & \(-1.354e-02\) & \(-2.429e-02\) & \(2.841e-01\) & \(2.583e-02\) & \(-2.728e-02\) & \(-2.091e+00\) & \(1.317e-01\) & \(4.010e+00\) & \(-2.763e+00\) \\
\(J0810 - r\) & \(2.849e-02\) & \(-1.158e-02\) & \(-6.615e-02\) & \(2.953e-01\) & \(2.854e-02\) & \(1.053e-01\) & \(-2.217e+00\) & \(3.368e-02\) & \(4.307e+00\) & \(-2.969e+00\) \\
\(J0820 - r\) & \(2.502e-02\) & \(-2.632e-02\) & \(-9.010e-02\) & \(3.023e-01\) & \(5.653e-02\) & \(1.871e-01\) & \(-2.283e+00\) & \(-3.066e-02\) & \(4.462e+00\) & \(-3.086e+00\) \\
\(J0830 - r\) & \(1.054e-03\) & \(-2.237e-02\) & \(-6.709e-02\) & \(2.889e-01\) & \(3.508e-02\) & \(1.075e-01\) & \(-2.174e+00\) & \(3.921e-02\) & \(4.197e+00\) & \(-2.899e+00\) \\
\(J0840 - r\) & \(-6.166e-03\) & \(-9.239e-03\) & \(-4.430e-02\) & \(2.853e-01\) & \(1.113e-02\) & \(3.689e-02\) & \(-2.112e+00\) & \(9.173e-02\) & \(4.013e+00\) & \(-2.757e+00\) \\
\(J0850 - r\) & \(-1.177e-02\) & \(5.265e-03\) & \(-3.267e-02\) & \(2.826e-01\) & \(-1.736e-02\) & \(3.247e-03\) & \(-2.069e+00\) & \(1.159e-01\) & \(3.875e+00\) & \(-2.637e+00\) \\
\(J0860 - r\) & \(-7.987e-03\) & \(-4.034e-03\) & \(-5.752e-02\) & \(2.881e-01\) & \(-1.355e-03\) & \(9.001e-02\) & \(-2.121e+00\) & \(4.659e-02\) & \(3.995e+00\) & \(-2.724e+00\) \\
\(J0870 - r\) & \(-1.296e-02\) & \(-2.692e-02\) & \(-1.086e-01\) & \(2.931e-01\) & \(3.293e-02\) & \(2.714e-01\) & \(-2.190e+00\) & \(-1.124e-01\) & \(4.175e+00\) & \(-2.869e+00\) \\
\(J0880 - r\) & \(-1.912e-02\) & \(-8.617e-03\) & \(-1.120e-01\) & \(2.985e-01\) & \(-4.045e-03\) & \(2.855e-01\) & \(-2.238e+00\) & \(-1.304e-01\) & \(4.284e+00\) & \(-2.971e+00\) \\
\(J0890 - r\) & \(-1.978e-02\) & \(-3.096e-04\) & \(-1.107e-01\) & \(2.925e-01\) & \(-2.751e-02\) & \(2.822e-01\) & \(-2.209e+00\) & \(-1.303e-01\) & \(4.230e+00\) & \(-2.952e+00\) \\
\(J0900 - r\) & \(6.038e-03\) & \(-1.397e-02\) & \(-1.237e-01\) & \(2.921e-01\) & \(2.835e-03\) & \(3.277e-01\) & \(-2.229e+00\) & \(-1.744e-01\) & \(4.295e+00\) & \(-3.022e+00\) \\
\(J0910 - r\) & \(-5.815e-03\) & \(-2.490e-02\) & \(-1.488e-01\) & \(2.969e-01\) & \(3.029e-02\) & \(4.070e-01\) & \(-2.277e+00\) & \(-2.267e-01\) & \(4.405e+00\) & \(-3.106e+00\) \\
\(J1007 - r\) & \(-6.468e-03\) & \(-2.092e-02\) & \(-1.233e-01\) & \(2.906e-01\) & \(3.044e-02\) & \(2.975e-01\) & \(-2.235e+00\) & \(-1.094e-01\) & \(4.262e+00\) & \(-2.985e+00\) \\
\(uJPAS - r\) & \(-1.473e-01\) & \(-1.094e-01\) & \(-3.239e-01\) & \(4.632e-01\) & \(9.431e-02\) & \(7.202e-01\) & \(-3.354e+00\) & \(-3.559e-02\) & \(7.902e+00\) & \(-2.168e+00\) \\
\(gSDSS - r\) & \(1.336e-02\) & \(-1.240e-03\) & \(-4.098e-02\) & \(2.275e-01\) & \(3.328e-03\) & \(5.943e-02\) & \(-1.695e+00\) & \(2.935e-02\) & \(4.271e+00\) & \(-2.266e+00\) \\
\(iSDSS - r\) & \(1.336e-02\) & \(-1.240e-03\) & \(-4.098e-02\) & \(2.275e-01\) & \(3.328e-03\) & \(5.943e-02\) & \(-1.695e+00\) & \(2.935e-02\) & \(3.271e+00\) & \(-2.266e+00\) \\
\(J - r\) & \(8.884e-03\) & \(-6.491e-02\) & \(-1.457e-01\) & \(2.926e-01\) & \(1.217e-01\) & \(3.479e-01\) & \(-2.280e+00\) & \(-1.267e-01\) & \(4.299e+00\) & \(-4.017e+00\) \\
\(H - r\) & \(6.271e-02\) & \(-1.715e-01\) & \(2.752e-02\) & \(1.974e-01\) & \(3.325e-01\) & \(-4.934e-01\) & \(-1.570e+00\) & \(7.894e-01\) & \(2.625e+00\) & \(-3.393e+00\) \\
\(K_s - r\) & \(3.755e-02\) & \(-1.461e-01\) & \(-7.684e-03\) & \(2.245e-01\) & \(2.795e-01\) & \(-3.336e-01\) & \(-1.734e+00\) & \(6.296e-01\) & \(2.844e+00\) & \(-3.570e+00\) \\
\(W1 - r\) & \(4.652e-03\) & \(-9.838e-02\) & \(-7.312e-02\) & \(2.459e-01\) & \(1.867e-01\) & \(-4.746e-02\) & \(-1.864e+00\) & \(3.509e-01\) & \(3.029e+00\) & \(-3.698e+00\) \\
\(W2 - r\) & \(-6.516e-02\) & \(-3.742e-03\) & \(-2.288e-01\) & \(2.962e-01\) & \(4.970e-03\) & \(6.106e-01\) & \(-2.231e+00\) & \(-2.167e-01\) & \(3.721e+00\) & \(-4.018e+00\) \\
\enddata

\tablenotetext{}{%
\footnotesize
Fitting coefficients for Stage\,II of the stellar loci. \(f(x, y) = a_{1} + a_{2} \cdot x + a_{3} \cdot y + a_{4} \cdot x^2 + a_{5} \cdot x \cdot y + a_{6} \cdot y^2 + a_{7} \cdot x^3 + a_{8} \cdot x^2 \cdot y + a_{9} \cdot y^2 \cdot x + a_{10} \cdot y^3\), where \(x \equiv g - i\) and \(y \equiv [\text{Fe/H}]\).
}

\end{deluxetable}

\begin{deluxetable}{llll}
\tabletypesize{\scriptsize}
\tablecaption{Extinction Coefficients ($E(B-V)=0.15$, $T_{\rm eff}=5500\,\mathrm{K}$, $\log g=4.5$)\label{tab:kx_ebv015_teff5500_logg45}}
\tablewidth{0pt}
\tablehead{
\colhead{Band } & \colhead{kx} & \colhead{Band } & \colhead{kx}
}
\startdata
$uJAVA$ & 4.8882 & $J0700$ & 2.1232 \\
$J0378$ & 4.6789 & $J0710$ & 2.0637 \\
$J0390$ & 4.5813 & $J0720$ & 2.0222 \\
$J0400$ & 4.4750 & $J0730$ & 1.9870 \\
$J0410$ & 4.3744 & $J0740$ & 1.9464 \\
$J0420$ & 4.2884 & $J0750$ & 1.9122 \\
$J0430$ & 4.1767 & $J0760$ & 1.8937 \\
$J0440$ & 4.0962 & $J0770$ & 1.8599 \\
$J0450$ & 3.9884 & $J0780$ & 1.7990 \\
$J0460$ & 3.8968 & $J0790$ & 1.7396 \\
$J0470$ & 3.8191 & $J0800$ & 1.6932 \\
$J0480$ & 3.7509 & $J0810$ & 1.6538 \\
$J0490$ & 3.6524 & $J0820$ & 1.6226 \\
$J0500$ & 3.5189 & $J0830$ & 1.5903 \\
$J0510$ & 3.4024 & $J0840$ & 1.5611 \\
$J0520$ & 3.2944 & $J0850$ & 1.5292 \\
$J0530$ & 3.2069 & $J0860$ & 1.4967 \\
$J0540$ & 3.1274 & $J0870$ & 1.4492 \\
$J0550$ & 2.9961 & $J0880$ & 1.4074 \\
$J0560$ & 2.8999 & $J0890$ & 1.3751 \\
$J0570$ & 2.8416 & $J0900$ & 1.3450 \\
$J0580$ & 2.7864 & $J0910$ & 1.3186 \\
$J0590$ & 2.7115 & $J1007$ & 1.1928 \\
$J0600$ & 2.6501 & $uJPAS$ & 4.7406 \\
$J0610$ & 2.5910 & $gSDSS$ & 3.6558 \\
$J0620$ & 2.5533 & $rSDSS$ & 2.5284 \\
$J0630$ & 2.5085 & $iSDSS$ & 1.8531 \\
$J0640$ & 2.4634 & $J$ & 0.7253 \\
$J0650$ & 2.3988 & $H$ & 0.4199 \\
$J0660$ & 2.3322 & $K_s$ & 0.2492 \\
$J0670$ & 2.2691 & $W1$ & 0.1081 \\
$J0680$ & 2.2212 & $W2$ & 0.0583 \\
$J0690$ & 2.1739 &  &  \\
\enddata
\tablenotetext{}{%
\footnotesize
Extinction coefficients for the miniJPAS, 2MASS, and WISE photometric bands.
}
\end{deluxetable}

\end{onecolumngrid} 

\clearpage
\bibliographystyle{aasjournalv7}
\bibliography{ref}

\end{CJK*}
\end{document}